\documentclass[aps,prl,floatfix,twocolumn,reprint,amsmath,amssymb,superscriptaddress,longbibliography]{revtex4-1}

\usepackage{amsfonts}
\usepackage{mathrsfs}
\usepackage{amsmath}
\usepackage{color}
\usepackage{graphicx}
\usepackage{bm}
\usepackage{amssymb}
\usepackage{xspace}
\usepackage{epstopdf}
\usepackage{dcolumn}
\usepackage{longtable}
\usepackage{multirow}
\usepackage{float}
\usepackage{comment}
\usepackage{soul}
\usepackage{makecell}
\usepackage{placeins}

\usepackage[colorlinks=true, letterpaper=true, pdfstartview=FitV, linkcolor=blue, citecolor=blue, urlcolor=blue]{hyperref}

\makeatother

\begin{document}

\title{Time-reversal-odd Transport in Odd-Parity Magnets}

\author{Ling Bai}
\affiliation{Key Lab of Advanced Optoelectronic Quantum Architecture and Measurement (MOE), Beijing Key Lab of Nanophotonics and Ultrafine Optoelectronic Systems, and School of Physics,Beijing Institute of Technology, Beijing 100081, China}

\author{Si Li}
\email{sili@nwu.edu.cn}
\affiliation{School of Physics, Northwest University, Xi'an 710127, China}

\author{Libor \v{S}mejkal}
\affiliation{Institute of Physics, Johannes Gutenberg University Mainz, 55099 Mainz, Germany}
\affiliation{Institute of Physics, Czech Academy of Sciences, Cukrovarnická 10, 162 00 Praha 6, Czech Republic}

\author{Yugui Yao}
\affiliation{Key Lab of Advanced Optoelectronic Quantum Architecture and Measurement (MOE), Beijing Key Lab of Nanophotonics and Ultrafine Optoelectronic Systems, and School of Physics,Beijing Institute of Technology, Beijing 100081, China}

\author{Wanxiang Feng}
\email{wxfeng@bit.edu.cn}
\affiliation{Key Lab of Advanced Optoelectronic Quantum Architecture and Measurement (MOE), Beijing Key Lab of Nanophotonics and Ultrafine Optoelectronic Systems, and School of Physics,Beijing Institute of Technology, Beijing 100081, China}

\date{\today}

\begin{abstract}
Unconventional magnets combining vanishing net magnetization with spin-split electronic structures and anomalous transport have attracted considerable attention. Using spin-group symmetry analysis, we derive general conditions for odd-parity spin splitting in two- and three-dimensional coplanar noncollinear antiferromagnets. We identify a key spin-group symmetry whose relativistic counterpart forbids time-reversal-odd transport and show that breaking it allows such transport in the presence of spin-orbit coupling. Tight-binding calculations validate these symmetry arguments and demonstrate an anomalous Hall effect in a fully compensated odd-parity magnetic state without an external field. By screening a three-dimensional magnetic materials database and constructing two-dimensional heterostructures, we further identify realistic platforms exhibiting pronounced spin-orientation-dependent anomalous Hall and magneto-optical effects. Our results establish a general symmetry framework for time-reversal-odd responses in odd-parity magnets and highlight fully compensated odd-parity magnets as a promising platform for controlling anomalous transport.
\end{abstract}

\maketitle

The recent emergence of unconventional compensated magnets has substantially broadened the landscape of magnetic materials beyond the conventional ferromagnet–antiferromagnet paradigm~\cite{2022PRXSmejkal1,2022PRXSmejkal2,2024AFMBai,2025NPLiu,2023arXivAnna,2025NatureYamada,2025NatureSong,2023NPtakagi,2024NatureZhu,2025PRLBai,2026NLBai}. A prominent example is altermagnetism, where compensated collinear magnetic order produces momentum-dependent nonrelativistic spin splitting and ferromagnetic-like responses despite vanishing net magnetization~\cite{2022PRXSmejkal1,2022PRXSmejkal2,2024AFMBai}. The associated spin splitting has even momentum parity and can exhibit $d$-, $g$-, or $i$-wave forms. This symmetry-based perspective has stimulated the exploration of other compensated magnetic phases, including odd-parity magnets with $p$- or $f$-wave spin textures~\cite{2023arXivAnna,2025NatureYamada,2025NatureSong}, noncoplanar antiferromagnets~\cite{2023NPtakagi,2024NatureZhu}, and type-IV collinear magnets~\cite{2025PRLBai,2026NLBai}. A central challenge is therefore to establish how distinct magnetic symmetries dictate spin-split electronic structures and, ultimately, experimentally observable responses.

Odd-parity magnets constitute a particularly intriguing class in which the spin polarization reverses under momentum inversion, $\bm{s}(\bm{k})=-\bm{s}(-\bm{k})$, with $p$-wave magnetism representing the lowest-order realization~\cite{2023arXivAnna,2024PRLBrekke,2024JPSCMaeda,2024PRBEzawa,2024PRBKudasov,2025arXivLuo,2025NatureYamada,2025NatureSong,2025NCChakraborty,2025PRLYuyue,2025PRBEzawa,2025PRBEzawa2,2025PRBSun,2025NewtonJungwirth,2025PRBPari,2025arXivOkumura,2025arXivLin,2025arXivZhou,2026PRXSong,2026PRBZeng,2026PRLHuang,2026PRLZhu,2026PRLLi,2026PRBliu,2026PRLFuPH,2026PRBDsouza,2026arXivMaxim,2026arXivLuo,2026arXivJohannes,2026arXivJan,2026arXivRobin,2026arXivZhang}. Such spin textures can emerge either from engineered mechanisms, including Floquet- and loop-current-induced states in collinear antiferromagnets~\cite{2026PRLHuang,2026PRLZhu,2026PRLLi,2026PRBliu}, or intrinsically in coplanar noncollinear magnets~\cite{2023arXivAnna,2025arXivLuo,2026PRXSong}. They offer promising routes toward spin filtering, tunneling magnetoresistance, nonequilibrium spin currents, and nonrelativistic Edelstein effects~\cite{2024PRLBrekke,2025NCChakraborty}. Experimentally, characteristic electronic anisotropy of $p$-wave magnetism has recently been detected through magnetotransport and photocurrent measurements~\cite{2025NatureYamada,2025NatureSong}. An anomalous Hall effect (AHE) has also been discussed in $p$-wave magnets, but existing realizations rely on slightly distorted magnetic configurations carrying a small auxiliary magnetization that breaks time-reversal symmetry~\cite{2025NatureYamada,2025arXivOkumura}. Whether a fully compensated odd-parity magnetic state can intrinsically generate time-reversal-odd transport such as the AHE without auxiliary magnetization or additional symmetry-lowering perturbations remains an open question. Resolving this issue requires a general symmetry framework that simultaneously determines the conditions for odd-parity spin splitting and the emergence of time-reversal-odd transport.

In this Letter, we demonstrate that fully compensated coplanar noncollinear antiferromagnets with odd-parity spin splitting can intrinsically host time-reversal-odd transport. We derive general symmetry criteria for odd-parity spin-momentum locking in two- and three-dimensional (2D and 3D) systems and identify the key symmetry condition that permits time-reversal-odd transport in the presence of spin-orbit coupling (SOC). Tight-binding calculations verify these criteria and show that SOC generates an intrinsic AHE in a fully compensated odd-parity magnetic state, without auxiliary magnetization or external symmetry breaking. Guided by the symmetry analysis, we identify candidates of 3D materials from the MAGNDATA database~\cite{2016ACGallego} and further construct a 2D heterostructure from experimentally realized monolayer building blocks. For the Fe$_3$C$_6$O$_6$–graphene–Fe$_3$C$_6$O$_6$ heterostructure, first-principles calculations reveal robust $f$-wave spin-momentum locking together with pronounced spin-orientation-dependent anomalous Hall and magneto-optical effects. These results establish a symmetry-based connection between odd-parity spin splitting and time-reversal-odd phenomena, revealing fully compensated odd-parity magnets as a distinct platform for electrically and optically detectable magnetic responses without net magnetization.

\begin{table*}[t!]
	\renewcommand{\arraystretch}{1.5}
	\caption{Symmetry criteria for realizing odd-parity spin splitting and time-reversal-odd transport in coplanar noncollinear magnets. At least one required symmetry must be present, while all forbidden symmetries must be absent.}
	\label{tab:Tab1}
	\begin{ruledtabular}
		\begin{tabular}{ccc}
			Symmetry criteria & 2D systems & 3D systems   \\ 
			\hline
			Required symmetries (at least one)    & {$[C_{\perp}(\theta) ||  E]$\ ($\theta \neq \pi$), $[C_{\perp}(\theta) || M_z]$} & $[C_{\perp}(\theta) ||  E]$\ ($\theta \neq \pi$) \\[6pt]
			Forbidden symmetries & \makecell{$[E \,\|\, P]$,$[C_{2\parallel} || E]$, $[C_{\perp}(\theta) || P]$,} & \makecell{$[E || P]$, $[C_{2\parallel} || E]$,} \\
			& \makecell{$[E || C_{2z}]$, $[C_{2\parallel} || M_z]$, $[C_{\perp}(\theta) || C_{2z}]$, $[C_{2\perp} || E]$} & \makecell{$[C_{\perp}(\theta) || P]$, $ [C_{2\perp} || E]$} 
		\end{tabular}
	\end{ruledtabular}
\end{table*}

{\emph{\textcolor{blue}{Symmetry analysis--}}}
For a coplanar, noncollinear magnetic system, the spin-only group can be written as
\begin{equation}
	\mathbf{r}_s=\{E,\bar{C}_{2\perp}\},
\end{equation}
where $E$ denotes the identity and $\bar C_{2\perp}=C_{2\perp}T$ represents a twofold spin rotation about an axis perpendicular to the spin plane, followed by time reversal. For a nondegenerate Bloch band, the spin-resolved energy can be written as $\varepsilon(s_{\perp},s_{\parallel},\bm{k})$, where $s_{\perp}$ and $s_{\parallel}$ denote the spin components perpendicular and parallel to the spin plane, respectively.
Under $\bar{C}_{2\perp}$, the energy band transforms as
\begin{equation}
	\bar{C}_{2\perp}\varepsilon(s_{\perp}, s_{\parallel}, \bm{k})=\varepsilon(-s_{\perp}, s_{\parallel}, -\bm{k}),
\end{equation}
in which the out-of-plane and in-plane spin components exhibit odd and even parity, respectively,
\begin{align}
	s_{\perp}(\bm{k})&=-s_{\perp}(-\bm{k}) \label{eq3}, \\
	s_{\parallel}(\bm{k})&=s_{\parallel}(-\bm{k}) \label{eq4}. 
\end{align}
To realize odd-parity spin splitting, the nontrivial spin-group operations must satisfy two conditions: (i) they must not impose an inversion-like transformation on $s_{\perp}$; and (ii) $s_{\parallel}$ must vanish for each nondegenerate spin-split band. In the following, we omit the translational parts of the spin-group operations, as they do not affect the spin-transformation constraints considered here.

For 3D systems, condition (i) excludes the symmetry operations $[E || P]$, $[C_{2\parallel} || E]$, and $[C_{\perp}(\theta)|| P]$. Here, $P$ denotes spatial inversion; $C_{2\parallel}$ denotes a twofold spin rotation about an axis lying in the spin plane; and $C_\perp(\theta)$ denotes a spin rotation by an arbitrary angle $\theta\neq 0\pmod{2\pi}$ about an axis perpendicular to the spin plane. Acting with $[E || P]$ and $[C_{\perp}(\theta) || P]$ on $s_{\perp}(\bm{k})$ yields, respectively,
\begin{align}
	[E || P]s_{\perp}(\bm{k})&=s_{\perp}(-\bm{k}), \\
	[C_{\perp}(\theta) || P]s_{\perp}(\bm{k})&=s_{\perp}(-\bm{k}).
\end{align}
Together with Eq.~\eqref{eq3}, these relations enforce $s_\perp(\mathbf{k})=-s_\perp(\mathbf{k})=0$, thereby forbidding odd-parity spin splitting. Moreover, applying $[C_{2\parallel} || E]$ to $s_{\perp}(\bm{k})$ gives
\begin{equation}
	[C_{2\parallel} || E]s_{\perp}(\bm{k})=-s_{\perp}(\bm{k}),
\end{equation}
sharing the same constraint. For 2D systems, the reduced dimensionality restricts the momentum to its in-plane component $k_{\parallel}$. Taking the $z$ axis as the out-of-plane direction, three additional symmetries, $[E || C_{2z}]$, $[C_{2\parallel} || M_z]$, and $[C_{\perp}(\theta) || C_{2z}]$, must be excluded, because $C_{2z}$ and $M_z$ play roles analogous to $P$ and $E$, i.e., $C_{2z} k_{\parallel} = -k_{\parallel}$ and $M_z k_{\parallel} = k_{\parallel}$, respectively. With these symmetry operations excluded, spin splitting is generally allowed at generic momenta throughout the Brillouin zone (BZ), except at specific high-symmetry points or lines where spin degeneracy is symmetry enforced, as well as at accidental band crossings.

For 3D systems, condition (ii) requires the symmetry $[C_{\perp}(\theta)||E]$, which directly enforces $s_{\parallel}=0$. For 2D systems, the symmetry $[C_{\perp}(\theta) || M_z]$ must also be imposed. Overall, under the above symmetry constraints, coplanar noncollinear magnets can retain the odd-parity spin texture of the out-of-plane component, while the in-plane component is strictly suppressed, thereby giving rise to odd-parity spin splitting throughout the band structure.

We next establish the symmetry conditions for time-reversal-odd transport in coplanar noncollinear odd-parity magnets. Taking the AHE as an example, it vanishes in the absence of SOC because the spin-only operation $\bar{C}_{2\perp}$ plays the role of an effective time-reversal symmetry. Thus SOC is necessary, but it still does not guarantee a finite AHE. Indeed, the SOC-free spin-group symmetry $[C_{\perp}(\theta)||E]$ may be present; for $\theta=\pi$ it becomes $[C_{2\perp}||E]$. When SOC locks spin to the lattice, this operation is identified with $T$ in the magnetic point group and consequently forbids the AHE. Therefore, the absence of $[C_{2\perp}||E]$ is a necessary symmetry requirement for realizing time-reversal-odd responses in odd-parity magnets. Table~\ref{tab:Tab1} summarizes the necessary spin-group symmetry conditions for the emergence of time-reversal-odd transport in coplanar noncollinear magnets with odd-parity spin splitting.

\begin{figure*}[htpb]
	\includegraphics[width=1.8\columnwidth]{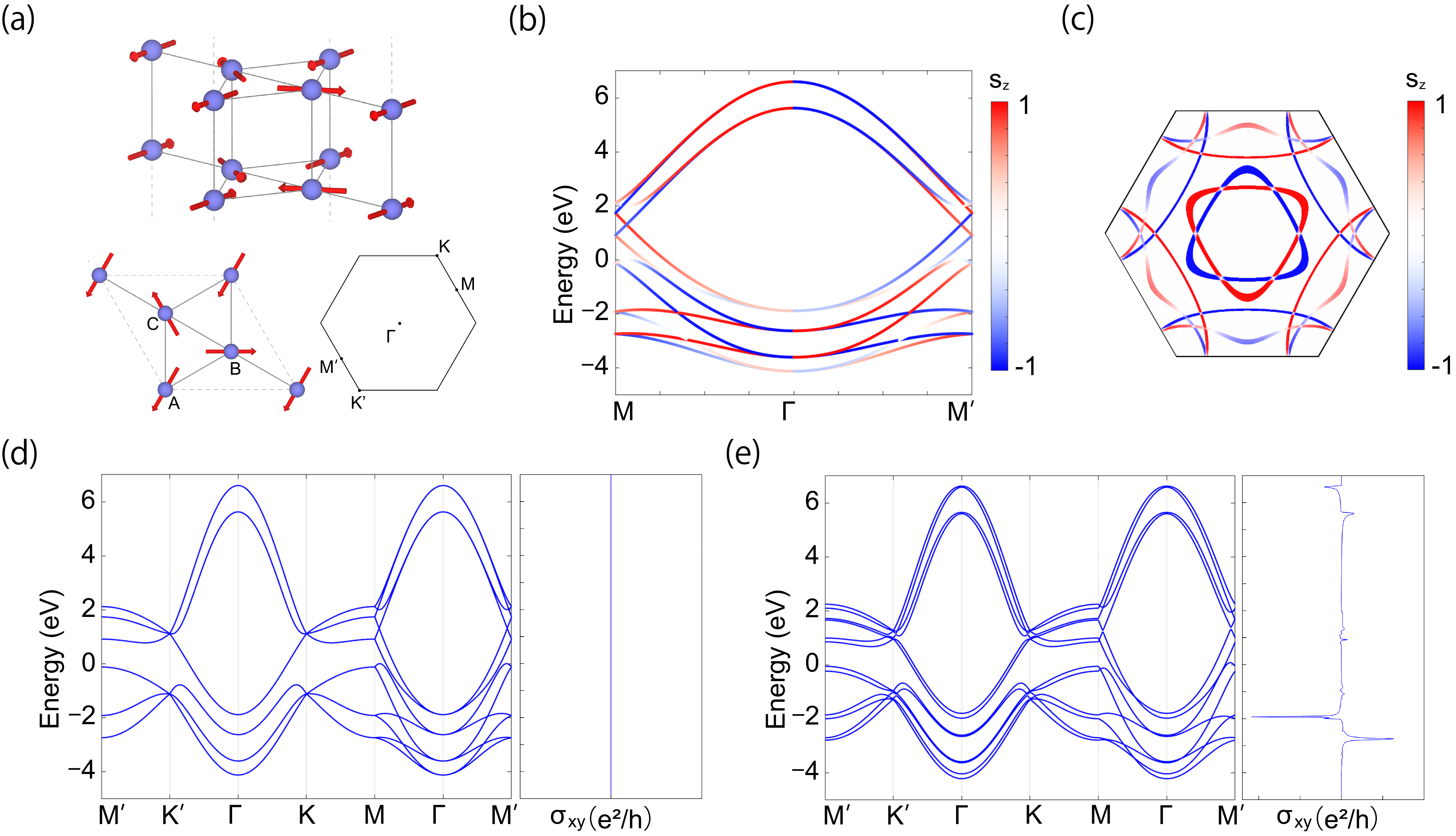}
	\caption{(a) Side and top views of the 2D bilayer $120^{\circ}$ antiferromagnetic order on a triangular lattice, together with the corresponding Brillouin zone. (b) Band structure along the M-$\Gamma$-M$^{\prime}$ path and (c) the isoenergy surface at $E=-2.1$ eV, characterized by the spin expectation $s_z(\mathbf{k})$. Band structures and anomalous Hall conductivities calculated using (d) SOC-free and (e) SOC-active models. The parameters are chosen as $t=J=1$ and $t_{\perp}=0.5$ in (d,e) and $r=\lambda_R=0.05$ in (e).}
	\label{fig1}
\end{figure*}

{\emph{\textcolor{blue}{Model--}}}
We consider a bilayer triangular-lattice system with coplanar noncollinear fully-compensated antiferromagnetic order, as schematically illustrated in Fig.~\ref{fig1}(a). Within each layer, the local magnetic moments form a 120° coplanar configuration on the triangular lattice, lying on the $xy$ plane. The unit cell contains six magnetic sublattices, (A$_t$, B$_t$, C$_t$, A$_b$, B$_b$, C$_b$). On the top layer, the magnetic vectors are $\mathbf{m}_{A_t} = (-1/2, -\sqrt{3}/2, 0)$, $\mathbf{m}_{B_t} = (1, 0, 0)$, and $\mathbf{m}_{C_t} = (-1/2, \sqrt{3}/2, 0)$, while the bottom layer hosts opposite moments, $\mathbf{m}_{A_b}=-\mathbf{m}_{A_t}$, $\mathbf{m}_{B_b}=-\mathbf{m}_{B_t}$, and $\mathbf{m}_{C_b}=-\mathbf{m}_{C_t}$. This bilayer system exhibits the symmetry $[C_z(2\pi/3)||E]$, which enforces $s_{x,y}(\bm{k})=0$. Meanwhile, it lacks the forbidden symmetries listed in Table~\ref{tab:Tab1}, ensuring that $s_z(\bm{k})=-s_z(-\bm{k})$. 

In the absence of SOC, we adopt the minimal tight-binding Hamiltonian~\cite{SuppMater}
\begin{align}
	{H}_0=& \sum_{\langle ij\rangle,\sigma} t \, {c}_{i\sigma}^\dagger \, {c}_{j\sigma}+\sum_{\langle ij\rangle,\sigma} t_{\perp} \, {c}_{i\sigma}^\dagger \, {c}_{j\sigma} + J \sum_{i, \sigma, \sigma'} \mathbf{m}_i \cdot c_{i \sigma}^\dagger \boldsymbol{\sigma}_{\sigma \sigma'} c_{i \sigma'}, \label{eq:H0}
\end{align}
where $c_{i\sigma}^{\dagger}$ ($c_{i\sigma}$) creates (annihilates) an electron at site $i$ with spin $\sigma$, and $\boldsymbol{\sigma}$ is the vector of Pauli matrices. The first (second) term describes intralayer (interlayer) nearest-neighbor hopping with amplitude $t$ ($t_\perp$), and the third term is the on-site exchange coupling to the noncollinear local moments $\mathbf{m}_i$ with strength $J$. Figure~\ref{fig1}(b) shows the $s_z$-projected bands along M-$\Gamma$-M$^\prime$, where a sizable spin splitting develops even without SOC; the split branches carry opposite $s_z$ under $\bm{k}\!\to\!-\bm{k}$, consistent with an odd-parity spin texture. The constant-energy contour at $E=-2.1$~eV in Fig.~\ref{fig1}(c) further corroborates this odd-parity structure and indicates an $f$-wave form factor. In Fig.~\ref{fig1}(d), we plot the band dispersions along several high-symmetry lines together with the AHC $\sigma_{xy}(E)$. Without SOC, the AHC vanishes identically ($\sigma_{xy}=0$) over the entire energy window, as required by the spin-only group symmetry $\bar{C}_{2z}$. This accounts for that despite the pronounced nonrelativistic spin splitting, the AHE remains forbidden in the SOC-free limit.

To analyze the SOC-induced transport signatures and the relevant symmetry-breaking mechanisms, we include the SOC-related term~\cite{SuppMater}
\begin{align}
	H_1=\sum_{\langle ij\rangle,\sigma,\sigma'} r \, c_{i\sigma}^\dagger \, c_{j\sigma'}
	+ i\lambda_R \sum_{\langle ij\rangle} c_i^\dagger \left[ (\boldsymbol{\sigma} \times \mathbf{d}_{ij}) \cdot \hat{\mathbf{z}} \right] c_j,
	\label{eq:H1}
\end{align}
where the first term denotes a spin-mixing interlayer hopping with amplitude $r$, and the second term is the Rashba SOC with coupling strength $\lambda_R$. Here, $\mathbf{d}_{ij}$ is the unit vector pointing from site $j$ to site $i$, and $\hat{\mathbf{z}}$ is the out-of-plane normal unit vector. The Rashba SOC arises from inversion-symmetry breaking at the bilayer interface. Figure~\ref{fig1}(e) presents the calculated band structures, together with the AHC obtained from the SOC-active Hamiltonian $H=H_0+H_1$. One can see that $\sigma_{xy}$ becomes finite over a broad energy window. This SOC-induced AHE can be understood as follows: the Rashba term and the spin-mixing hopping break the symmetries (e.g. $\bar{C}_{2z}$) that suppress the Hall response, thereby allowing a net Berry curvature to accumulate over the BZ and yielding a nonzero $\sigma_{xy}$.

\begin{figure}
	\centering
	\includegraphics[width=1\columnwidth]{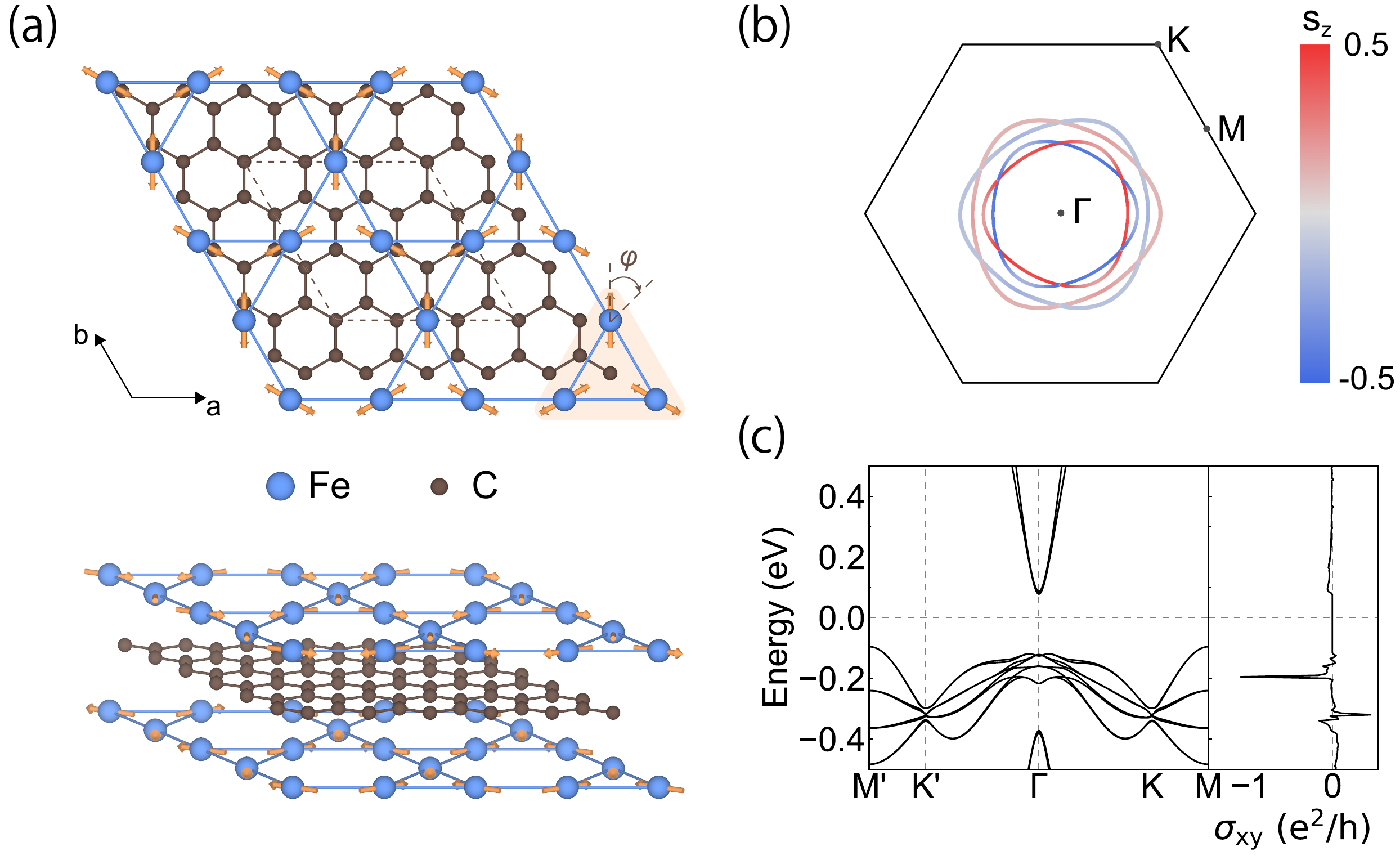}
	\caption{(a) Top and side views of crystal and magnetic structures of the Fe$_3$C$_6$O$_6$-graphene-Fe$_3$C$_6$O$_6$ (Fe-C-Fe) heterostructure. For clarity, the C and O atoms in the Fe$_3$C$_6$O$_6$ layers are omitted. Gold arrows denote the spin magnetic moments on Fe atoms, forming a $120^\circ$ coplanar noncollinear configuration. The angle $\varphi$ characterizes a uniform in-plane rotation of the three spins. (b) $s_z$-resolved nonrelativistic Fermi surface at an energy 1.04 eV below the valence-band maximum. (c) Relativistic band structure (left) and anomalous Hall conductivity $\sigma_{xy}$ at $\varphi=0^\circ$ (right).}
	\label{fig2}
\end{figure}

{\emph{\textcolor{blue}{Material realization--}}}
Guided by the above symmetry analysis, we identify candidates of odd-parity magnetism with time-reversal-odd transport by screening the MAGNDATA database of 3D magnetic materials~\cite{2016ACGallego} and constructing 2D heterostructures from experimentally realized monolayer materials. We here focus on the 2D case, while the 3D candidates and the corresponding results are included in Supplemental Table~S1 and Fig. S1~\cite{SuppMater}.

As a representative 2D platform, we build a Fe$_3$C$_6$O$_6$-graphene-Fe$_3$C$_6$O$_6$ (Fe-C-Fe) heterostructure, as shown in Fig.~\ref{fig2}(a). The parent material Fe$_3$C$_6$O$_6$ is a kagome metal-organic framework whose magnetic ground state hosts a coplanar $120^\circ$ noncollinear spin texture (Supplemental Fig.~S2~\cite{SuppMater}) with positive vector spin chirality $\kappa=+1$~\cite{2023NLZhou}. Such coplanar textures can, in principle, be reoriented by external fields or spin torques, as demonstrated in related triangular noncollinear antiferromagnets~\cite{2016SANayak,2021NMTakeuchi}. We therefore parametrize the in-plane spin orientation by a global rotation angle $\varphi$, while preserving the mutual $120^\circ$ arrangement of the three Fe moments. In monolayer Fe$_3$C$_6$O$_6$, the spin-group operation $[E||P]$ is preserved for arbitrary $\varphi$, thereby forbidding odd-parity spin splitting. To realize odd-parity antiferromagnetism, we insert a graphene layer between the two layers of an AA-stacked Fe$_3$C$_6$O$_6$ bilayer to form the Fe-C-Fe heterostructure. The two Fe$_3$C$_6$O$_6$ layers couple antiferromagnetically, with corresponding Fe moments aligned antiparallel and a fully compensated net magnetization. Because the inversion centers of graphene and the Fe$_3$C$_6$O$_6$ bilayer are spatially offset, the heterostructure is noncentrosymmetric and no longer possesses the spin-group operation $[E||P]$. For $\varphi=0^\circ$, its spin space group is $p^{6^1_{001}}\bar{6}^{2_{120}}m^{2_{010}}2^{m_{001}}1$~\cite{2026arXivYu}, which allows odd-parity spin splitting (Supplemental Table~S2~\cite{SuppMater}). Specifically, this spin space group contains the operation $[C_{2[100]}\,\|\,C_{2[120]}]$, which imposes
\begin{equation}
	[C_{2[100]}\,\|\, C_{2[120]}]\varepsilon(s_z,k_x,k_y)=\varepsilon(-s_z,-k_x,k_y),
\end{equation}
enforcing spin degeneracy along the $k_y$ axis. Owing to the threefold rotation symmetry $C_{3z}$, two additional symmetry-equivalent spin-degenerate lines appear at $k_x=\pm\sqrt{3}k_y$. Together, these symmetries produce an $f$-wave spin-momentum-locking pattern, as shown in Fig.~\ref{fig2}(b) and Supplemental Fig.~S3~\cite{SuppMater}.

With SOC included, the magnetic space group at $\varphi=0^\circ$ becomes $p\bar{6}m^\prime2^\prime$, which permits an AHE. Figure~\ref{fig2}(c) shows the relativistic band structure and the energy-dependent AHC $\sigma_{xy}$. While $\sigma_{xy}$ vanishes at charge neutrality, it reaches $-1.11~e^2/h$ ($0.46~e^2/h$) at $0.10$ ($0.23$) eV below the valence-band maximum under hole doping. The magnitude drops rapidly away from these energies, indicating that the anomalous Hall response is dominated by a few Berry-curvature hot spots rather than being uniformly distributed across the valence bands.

We next examine the spin space group, magnetic space group, odd-parity spin-momentum-locking type, and symmetry-allowance of AHE for different $\varphi$ (Supplemental Table S2~\cite{SuppMater}). The $f$-wave spin-momentum locking persists throughout the rotation because the symmetries $[C_{2\parallel}|C_{2[120]}]$ and $[C_{3z}^{+}||C_{3z}^{+}]$ are preserved for all $\varphi$. Among the considered spin orientations, the AHE is symmetry-forbidden only at $\varphi=90^\circ$. At this angle, the magnetic space group becomes $p\bar{6}m2$, which contains the symmetries $C_{2\parallel}$ and $M_{\parallel}$, either of which enforces $\sigma_{xy}=0$ and hence forbids the AHE.

\begin{figure}
	\centering
	\includegraphics[width=1\columnwidth]{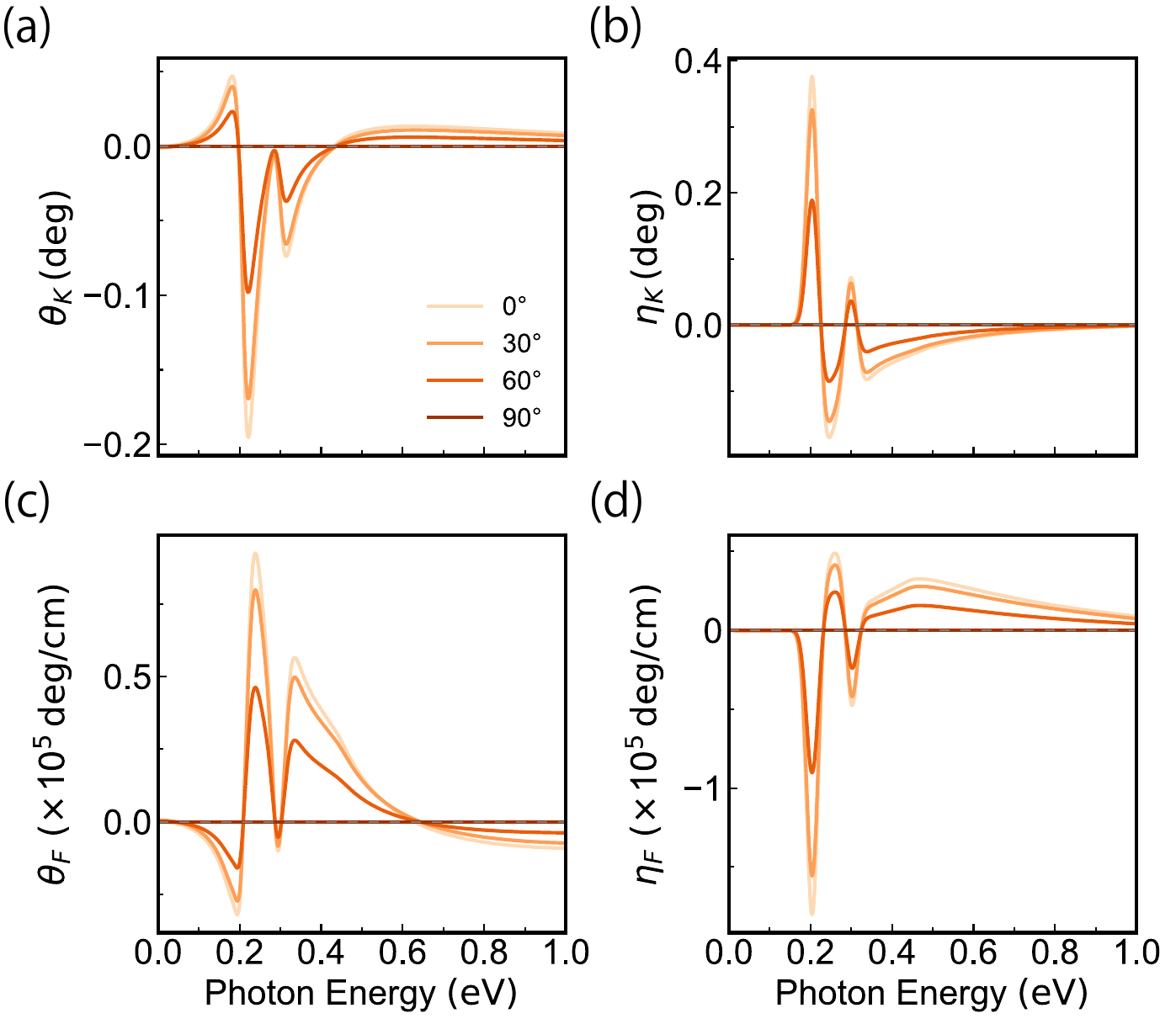}
	\caption{(a,b) Kerr rotation angle and ellipticity and (c,d) Faraday  rotation angle and ellipticity of Fe-C-Fe heterostructure for $\varphi=0^\circ$, $30^\circ$, $60^\circ$, and $90^\circ$.}
	\label{fig3}
\end{figure}

Motivated by the close connection between magneto-optical effects and the AHE (the latter corresponding to the $\omega\!\to\!0$ limit of the off-diagonal optical conductivity $\sigma_{xy}(\omega)$), we investigate spin-orientation-dependent Kerr and Faraday effects in the Fe-C-Fe heterostructure. The calculated optical conductivities for $\varphi=0^\circ$, $30^\circ$, $60^\circ$, and $90^\circ$ are shown in Supplemental Fig.~S5~\cite{SuppMater}. For all four spin orientations, the spectra of both the real and imaginary parts of the longitudinal conductivity, $\sigma_{xx}'$ and $\sigma_{xx}''$, are nearly identical, consistent with previous reports on Mn$_3$\textit{X} and Mn$_3$\textit{X}N compounds~\cite{2015PRBFeng,2019PRBZhou}. In contrast, the transverse optical conductivity $\sigma_{xy}(\omega)$ depends strongly on $\varphi$: its overall magnitude decreases with increasing $\varphi$ and vanishes at $\varphi=90^\circ$. The resulting Kerr and Faraday spectra are shown in Fig.~\ref{fig3}. As $\varphi$ increases, the Kerr/Faraday rotation angles $\theta_{\mathrm{K/F}}$ and ellipticities $\eta_{\mathrm{K/F}}$ are suppressed and become zero at $\varphi=90^\circ$, tracking the evolution of $\sigma_{xy}(\omega)$. The magneto-optical response is maximized at $\varphi=0^\circ$, where the Kerr and Faraday rotation angles reach $0.195^\circ$ and $0.92\times10^{5}$ deg/cm, respectively; the Kerr rotation is comparable in magnitude to that reported for other 2D ferromagnetic materials~\cite{2017NatureHuang,2017NanoSZhou}. These results establish the spin-orientation angle as an effective control knob for tuning the magneto-optical responses of odd-parity magnets.

{\emph{\textcolor{blue}{Conclusion.--}}}
In summary, combining spin group and magnetic group symmetry analyses, we established general criteria for realizing odd-parity spin splitting in coplanar noncollinear antiferromagnets and for enabling time-reversal-odd transport in the presence of SOC. In particular, we identified the symmetries that, once promoted to the relativistic setting, forbid the intrinsic AHE; breaking them allows a finite Hall response even in a fully compensated state with zero net magnetization. Tight-binding calculations corroborate these symmetry arguments and demonstrate an SOC-induced AHE without auxiliary magnetization or external symmetry breaking. Guided by these symmetry rules, database screening and first-principles calculations further pinpoint realistic platforms that host robust odd-parity spin-momentum locking and pronounced, spin-orientation-tunable anomalous Hall and magneto-optical effects. These results provide a symmetry-based route to engineering electrically and optically detectable time-reversal-odd responses in fully compensated odd-parity magnets.

{\emph{\textcolor{blue}{Acknowledgments.--}}}
This work is supported by the National Natural Science Foundation of China (Grants W2511003, 12274027, 12234003, and 12321004), the National Key R\&D Program of China (Grants 2022YFA1402600 and 2022YFA1403800), the Fundamental Research Funds for the Central Universities (Grant 2024CX06104), and the Key Program (Grant No. 2025JC-QYCX-007) and the Youth Project (Category B) (Grant No. 2026JC-YXQN-026) of the Natural Science Basic Research Plan of Shaanxi Province.

\bibliographystyle{apsrev4-2}
\bibliography{mybib}

\begin{thebibliography}{64}%
\makeatletter
\providecommand \@ifxundefined [1]{%
 \@ifx{#1\undefined}
}%
\providecommand \@ifnum [1]{%
 \ifnum #1\expandafter \@firstoftwo
 \else \expandafter \@secondoftwo
 \fi
}%
\providecommand \@ifx [1]{%
 \ifx #1\expandafter \@firstoftwo
 \else \expandafter \@secondoftwo
 \fi
}%
\providecommand \natexlab [1]{#1}%
\providecommand \enquote  [1]{``#1''}%
\providecommand \bibnamefont  [1]{#1}%
\providecommand \bibfnamefont [1]{#1}%
\providecommand \citenamefont [1]{#1}%
\providecommand \href@noop [0]{\@secondoftwo}%
\providecommand \href [0]{\begingroup \@sanitize@url \@href}%
\providecommand \@href[1]{\@@startlink{#1}\@@href}%
\providecommand \@@href[1]{\endgroup#1\@@endlink}%
\providecommand \@sanitize@url [0]{\catcode `\\12\catcode `\$12\catcode
  `\&12\catcode `\#12\catcode `\^12\catcode `\_12\catcode `\%12\relax}%
\providecommand \@@startlink[1]{}%
\providecommand \@@endlink[0]{}%
\providecommand \url  [0]{\begingroup\@sanitize@url \@url }%
\providecommand \@url [1]{\endgroup\@href {#1}{\urlprefix }}%
\providecommand \urlprefix  [0]{URL }%
\providecommand \Eprint [0]{\href }%
\providecommand \doibase [0]{https://doi.org/}%
\providecommand \selectlanguage [0]{\@gobble}%
\providecommand \bibinfo  [0]{\@secondoftwo}%
\providecommand \bibfield  [0]{\@secondoftwo}%
\providecommand \translation [1]{[#1]}%
\providecommand \BibitemOpen [0]{}%
\providecommand \bibitemStop [0]{}%
\providecommand \bibitemNoStop [0]{.\EOS\space}%
\providecommand \EOS [0]{\spacefactor3000\relax}%
\providecommand \BibitemShut  [1]{\csname bibitem#1\endcsname}%
\let\auto@bib@innerbib\@empty
\bibitem [{\citenamefont {\ifmmode~\check{S}\else \v{S}\fi{}mejkal}\ \emph
  {et~al.}(2022{\natexlab{a}})\citenamefont {\ifmmode~\check{S}\else
  \v{S}\fi{}mejkal}, \citenamefont {Sinova},\ and\ \citenamefont
  {Jungwirth}}]{2022PRXSmejkal1}%
  \BibitemOpen
  \bibfield  {author} {\bibinfo {author} {\bibfnamefont {L.}~\bibnamefont
  {\ifmmode~\check{S}\else \v{S}\fi{}mejkal}}, \bibinfo {author} {\bibfnamefont
  {J.}~\bibnamefont {Sinova}},\ and\ \bibinfo {author} {\bibfnamefont
  {T.}~\bibnamefont {Jungwirth}},\ }\bibfield  {title} {\bibinfo {title}
  {Beyond conventional ferromagnetism and antiferromagnetism: A phase with
  nonrelativistic spin and crystal rotation symmetry},\ }\href
  {https://doi.org/10.1103/PhysRevX.12.031042} {\bibfield  {journal} {\bibinfo
  {journal} {Phys. Rev. X}\ }\textbf {\bibinfo {volume} {12}},\ \bibinfo
  {pages} {031042} (\bibinfo {year} {2022}{\natexlab{a}})}\BibitemShut
  {NoStop}%
\bibitem [{\citenamefont {\ifmmode~\check{S}\else \v{S}\fi{}mejkal}\ \emph
  {et~al.}(2022{\natexlab{b}})\citenamefont {\ifmmode~\check{S}\else
  \v{S}\fi{}mejkal}, \citenamefont {Sinova},\ and\ \citenamefont
  {Jungwirth}}]{2022PRXSmejkal2}%
  \BibitemOpen
  \bibfield  {author} {\bibinfo {author} {\bibfnamefont {L.}~\bibnamefont
  {\ifmmode~\check{S}\else \v{S}\fi{}mejkal}}, \bibinfo {author} {\bibfnamefont
  {J.}~\bibnamefont {Sinova}},\ and\ \bibinfo {author} {\bibfnamefont
  {T.}~\bibnamefont {Jungwirth}},\ }\bibfield  {title} {\bibinfo {title}
  {Emerging research landscape of altermagnetism},\ }\href
  {https://doi.org/10.1103/PhysRevX.12.040501} {\bibfield  {journal} {\bibinfo
  {journal} {Phys. Rev. X}\ }\textbf {\bibinfo {volume} {12}},\ \bibinfo
  {pages} {040501} (\bibinfo {year} {2022}{\natexlab{b}})}\BibitemShut
  {NoStop}%
\bibitem [{\citenamefont {Bai}\ \emph {et~al.}(2024)\citenamefont {Bai},
  \citenamefont {Feng}, \citenamefont {Liu}, \citenamefont {{\v S}mejkal},
  \citenamefont {Mokrousov},\ and\ \citenamefont {Yao}}]{2024AFMBai}%
  \BibitemOpen
  \bibfield  {author} {\bibinfo {author} {\bibfnamefont {L.}~\bibnamefont
  {Bai}}, \bibinfo {author} {\bibfnamefont {W.}~\bibnamefont {Feng}}, \bibinfo
  {author} {\bibfnamefont {S.}~\bibnamefont {Liu}}, \bibinfo {author}
  {\bibfnamefont {L.}~\bibnamefont {{\v S}mejkal}}, \bibinfo {author}
  {\bibfnamefont {Y.}~\bibnamefont {Mokrousov}},\ and\ \bibinfo {author}
  {\bibfnamefont {Y.}~\bibnamefont {Yao}},\ }\bibfield  {title} {\bibinfo
  {title} {Altermagnetism: Exploring new frontiers in magnetism and
  spintronics},\ }\href {https://doi.org/10.1002/adfm.202409327} {\bibfield
  {journal} {\bibinfo  {journal} {Adv. Funct. Mater.}\ }\textbf {\bibinfo
  {volume} {34}},\ \bibinfo {pages} {2409327} (\bibinfo {year}
  {2024})}\BibitemShut {NoStop}%
\bibitem [{\citenamefont {Liu}\ \emph {et~al.}(2025)\citenamefont {Liu},
  \citenamefont {Dai},\ and\ \citenamefont {Bl{\"u}gel}}]{2025NPLiu}%
  \BibitemOpen
  \bibfield  {author} {\bibinfo {author} {\bibfnamefont {Q.}~\bibnamefont
  {Liu}}, \bibinfo {author} {\bibfnamefont {X.}~\bibnamefont {Dai}},\ and\
  \bibinfo {author} {\bibfnamefont {S.}~\bibnamefont {Bl{\"u}gel}},\ }\bibfield
   {title} {\bibinfo {title} {Different facets of unconventional magnetism},\
  }\href {https://doi.org/10.1038/s41567-024-02750-3} {\bibfield  {journal}
  {\bibinfo  {journal} {Nat. Phys.}\ }\textbf {\bibinfo {volume} {21}},\
  \bibinfo {pages} {329} (\bibinfo {year} {2025})}\BibitemShut {NoStop}%
\bibitem [{\citenamefont {Hellenes}\ \emph {et~al.}(2023)\citenamefont
  {Hellenes}, \citenamefont {Jungwirth}, \citenamefont {Jaeschke-Ubiergo},
  \citenamefont {Chakraborty}, \citenamefont {Sinova},\ and\ \citenamefont
  {Šmejkal}}]{2023arXivAnna}%
  \BibitemOpen
  \bibfield  {author} {\bibinfo {author} {\bibfnamefont {A.~B.}\ \bibnamefont
  {Hellenes}}, \bibinfo {author} {\bibfnamefont {T.}~\bibnamefont {Jungwirth}},
  \bibinfo {author} {\bibfnamefont {R.}~\bibnamefont {Jaeschke-Ubiergo}},
  \bibinfo {author} {\bibfnamefont {A.}~\bibnamefont {Chakraborty}}, \bibinfo
  {author} {\bibfnamefont {J.}~\bibnamefont {Sinova}},\ and\ \bibinfo {author}
  {\bibfnamefont {L.}~\bibnamefont {Šmejkal}},\ }\bibfield  {title} {\bibinfo
  {title} {P-wave magnets},\ }\href {https://arxiv.org/abs/2309.01607}
  {\bibfield  {journal} {\bibinfo  {journal} {arXiv:2309.01607}\ } (\bibinfo
  {year} {2023})}\BibitemShut {NoStop}%
\bibitem [{\citenamefont {Yamada}\ \emph {et~al.}(2025)\citenamefont {Yamada}
  \emph {et~al.}}]{2025NatureYamada}%
  \BibitemOpen
  \bibfield  {author} {\bibinfo {author} {\bibfnamefont {R.}~\bibnamefont
  {Yamada}} \emph {et~al.},\ }\bibfield  {title} {\bibinfo {title} {A metallic
  $p$-wave magnet with commensurate spin helix},\ }\href
  {https://doi.org/10.1038/s41586-025-09404-3} {\bibfield  {journal} {\bibinfo
  {journal} {Nature}\ }\textbf {\bibinfo {volume} {646}},\ \bibinfo {pages}
  {837} (\bibinfo {year} {2025})}\BibitemShut {NoStop}%
\bibitem [{\citenamefont {Song}\ \emph {et~al.}(2025)\citenamefont {Song},
  \citenamefont {Stavri{\'c}}, \citenamefont {Barone}, \citenamefont
  {Droghetti}, \citenamefont {Antonenko}, \citenamefont {Venderbos},
  \citenamefont {Occhialini}, \citenamefont {Ilyas}, \citenamefont
  {Erge{\c{c}}en}, \citenamefont {Gedik}, \citenamefont {Cheong}, \citenamefont
  {Fernandes}, \citenamefont {Picozzi},\ and\ \citenamefont
  {Comin}}]{2025NatureSong}%
  \BibitemOpen
  \bibfield  {author} {\bibinfo {author} {\bibfnamefont {Q.}~\bibnamefont
  {Song}}, \bibinfo {author} {\bibfnamefont {S.}~\bibnamefont {Stavri{\'c}}},
  \bibinfo {author} {\bibfnamefont {P.}~\bibnamefont {Barone}}, \bibinfo
  {author} {\bibfnamefont {A.}~\bibnamefont {Droghetti}}, \bibinfo {author}
  {\bibfnamefont {D.~S.}\ \bibnamefont {Antonenko}}, \bibinfo {author}
  {\bibfnamefont {J.~W.}\ \bibnamefont {Venderbos}}, \bibinfo {author}
  {\bibfnamefont {C.~A.}\ \bibnamefont {Occhialini}}, \bibinfo {author}
  {\bibfnamefont {B.}~\bibnamefont {Ilyas}}, \bibinfo {author} {\bibfnamefont
  {E.}~\bibnamefont {Erge{\c{c}}en}}, \bibinfo {author} {\bibfnamefont
  {N.}~\bibnamefont {Gedik}}, \bibinfo {author} {\bibfnamefont {S.-W.}\
  \bibnamefont {Cheong}}, \bibinfo {author} {\bibfnamefont {R.~M.}\
  \bibnamefont {Fernandes}}, \bibinfo {author} {\bibfnamefont {S.}~\bibnamefont
  {Picozzi}},\ and\ \bibinfo {author} {\bibfnamefont {R.}~\bibnamefont
  {Comin}},\ }\bibfield  {title} {\bibinfo {title} {Electrical switching of a
  $p$-wave magnet},\ }\href {https://doi.org/10.1038/s41586-025-08821-6}
  {\bibfield  {journal} {\bibinfo  {journal} {Nature}\ }\textbf {\bibinfo
  {volume} {642}},\ \bibinfo {pages} {64} (\bibinfo {year} {2025})}\BibitemShut
  {NoStop}%
\bibitem [{\citenamefont {Takagi}\ \emph {et~al.}(2023)\citenamefont {Takagi},
  \citenamefont {Takagi}, \citenamefont {Minami}, \citenamefont {Nomoto},
  \citenamefont {Ohishi}, \citenamefont {Suzuki}, \citenamefont {Yanagi},
  \citenamefont {Hirayama}, \citenamefont {Khanh}, \citenamefont {Karube},
  \citenamefont {Saito}, \citenamefont {Hashizume}, \citenamefont {Kiyanagi},
  \citenamefont {Tokura}, \citenamefont {Arita}, \citenamefont {Nakajima},\
  and\ \citenamefont {Seki}}]{2023NPtakagi}%
  \BibitemOpen
  \bibfield  {author} {\bibinfo {author} {\bibfnamefont {H.}~\bibnamefont
  {Takagi}}, \bibinfo {author} {\bibfnamefont {R.}~\bibnamefont {Takagi}},
  \bibinfo {author} {\bibfnamefont {S.}~\bibnamefont {Minami}}, \bibinfo
  {author} {\bibfnamefont {T.}~\bibnamefont {Nomoto}}, \bibinfo {author}
  {\bibfnamefont {K.}~\bibnamefont {Ohishi}}, \bibinfo {author} {\bibfnamefont
  {M.-T.}\ \bibnamefont {Suzuki}}, \bibinfo {author} {\bibfnamefont
  {Y.}~\bibnamefont {Yanagi}}, \bibinfo {author} {\bibfnamefont
  {M.}~\bibnamefont {Hirayama}}, \bibinfo {author} {\bibfnamefont {N.~D.}\
  \bibnamefont {Khanh}}, \bibinfo {author} {\bibfnamefont {K.}~\bibnamefont
  {Karube}}, \bibinfo {author} {\bibfnamefont {H.}~\bibnamefont {Saito}},
  \bibinfo {author} {\bibfnamefont {D.}~\bibnamefont {Hashizume}}, \bibinfo
  {author} {\bibfnamefont {R.}~\bibnamefont {Kiyanagi}}, \bibinfo {author}
  {\bibfnamefont {Y.}~\bibnamefont {Tokura}}, \bibinfo {author} {\bibfnamefont
  {R.}~\bibnamefont {Arita}}, \bibinfo {author} {\bibfnamefont
  {T.}~\bibnamefont {Nakajima}},\ and\ \bibinfo {author} {\bibfnamefont
  {S.}~\bibnamefont {Seki}},\ }\bibfield  {title} {\bibinfo {title}
  {Spontaneous topological {Hall} effect induced by non-coplanar
  antiferromagnetic order in intercalated van der {Waals} materials},\ }\href
  {https://doi.org/10.1038/s41567-023-02017-3} {\bibfield  {journal} {\bibinfo
  {journal} {Nat. Phys.}\ }\textbf {\bibinfo {volume} {19}},\ \bibinfo {pages}
  {961} (\bibinfo {year} {2023})}\BibitemShut {NoStop}%
\bibitem [{\citenamefont {Zhu}\ \emph {et~al.}(2024)\citenamefont {Zhu} \emph
  {et~al.}}]{2024NatureZhu}%
  \BibitemOpen
  \bibfield  {author} {\bibinfo {author} {\bibfnamefont {Y.-P.}\ \bibnamefont
  {Zhu}} \emph {et~al.},\ }\bibfield  {title} {\bibinfo {title} {Observation of
  plaid-like spin splitting in a noncoplanar antiferromagnet},\ }\href
  {https://doi.org/10.1038/s41586-024-07023-w} {\bibfield  {journal} {\bibinfo
  {journal} {Nature}\ }\textbf {\bibinfo {volume} {626}},\ \bibinfo {pages}
  {523} (\bibinfo {year} {2024})}\BibitemShut {NoStop}%
\bibitem [{\citenamefont {Bai}\ \emph {et~al.}(2025)\citenamefont {Bai},
  \citenamefont {Zhang}, \citenamefont {Feng},\ and\ \citenamefont
  {Yao}}]{2025PRLBai}%
  \BibitemOpen
  \bibfield  {author} {\bibinfo {author} {\bibfnamefont {L.}~\bibnamefont
  {Bai}}, \bibinfo {author} {\bibfnamefont {R.-W.}\ \bibnamefont {Zhang}},
  \bibinfo {author} {\bibfnamefont {W.}~\bibnamefont {Feng}},\ and\ \bibinfo
  {author} {\bibfnamefont {Y.}~\bibnamefont {Yao}},\ }\bibfield  {title}
  {\bibinfo {title} {Anomalous {Hall} effect in type {IV} {2D} collinear
  magnets},\ }\href {https://doi.org/10.1103/PhysRevLett.135.036702} {\bibfield
   {journal} {\bibinfo  {journal} {Phys. Rev. Lett.}\ }\textbf {\bibinfo
  {volume} {135}},\ \bibinfo {pages} {036702} (\bibinfo {year}
  {2025})}\BibitemShut {NoStop}%
\bibitem [{\citenamefont {Bai}\ \emph {et~al.}(2026)\citenamefont {Bai},
  \citenamefont {Liu}, \citenamefont {Wang}, \citenamefont {Šmejkal},
  \citenamefont {Sinova}, \citenamefont {Mokrousov}, \citenamefont {Yao},\ and\
  \citenamefont {Feng}}]{2026NLBai}%
  \BibitemOpen
  \bibfield  {author} {\bibinfo {author} {\bibfnamefont {L.}~\bibnamefont
  {Bai}}, \bibinfo {author} {\bibfnamefont {S.}~\bibnamefont {Liu}}, \bibinfo
  {author} {\bibfnamefont {X.}~\bibnamefont {Wang}}, \bibinfo {author}
  {\bibfnamefont {L.}~\bibnamefont {Šmejkal}}, \bibinfo {author}
  {\bibfnamefont {J.}~\bibnamefont {Sinova}}, \bibinfo {author} {\bibfnamefont
  {Y.}~\bibnamefont {Mokrousov}}, \bibinfo {author} {\bibfnamefont
  {Y.}~\bibnamefont {Yao}},\ and\ \bibinfo {author} {\bibfnamefont
  {W.}~\bibnamefont {Feng}},\ }\bibfield  {title} {\bibinfo {title}
  {{PT}-symmetric antiferromagnets as building blocks for anomalous
  transport},\ }\href {https://doi.org/10.1021/acs.nanolett.6c00290} {\bibfield
   {journal} {\bibinfo  {journal} {Nano Lett.}\ }\textbf {\bibinfo {volume}
  {26}},\ \bibinfo {pages} {3934} (\bibinfo {year} {2026})}\BibitemShut
  {NoStop}%
\bibitem [{\citenamefont {Brekke}\ \emph {et~al.}(2024)\citenamefont {Brekke},
  \citenamefont {Sukhachov}, \citenamefont {Giil}, \citenamefont {Brataas},\
  and\ \citenamefont {Linder}}]{2024PRLBrekke}%
  \BibitemOpen
  \bibfield  {author} {\bibinfo {author} {\bibfnamefont {B.}~\bibnamefont
  {Brekke}}, \bibinfo {author} {\bibfnamefont {P.}~\bibnamefont {Sukhachov}},
  \bibinfo {author} {\bibfnamefont {H.~G.}\ \bibnamefont {Giil}}, \bibinfo
  {author} {\bibfnamefont {A.}~\bibnamefont {Brataas}},\ and\ \bibinfo {author}
  {\bibfnamefont {J.}~\bibnamefont {Linder}},\ }\bibfield  {title} {\bibinfo
  {title} {Minimal models and transport properties of unconventional $p$-wave
  magnets},\ }\href {https://doi.org/10.1103/PhysRevLett.133.236703} {\bibfield
   {journal} {\bibinfo  {journal} {Phys. Rev. Lett.}\ }\textbf {\bibinfo
  {volume} {133}},\ \bibinfo {pages} {236703} (\bibinfo {year}
  {2024})}\BibitemShut {NoStop}%
\bibitem [{\citenamefont {Maeda}\ \emph {et~al.}(2024)\citenamefont {Maeda},
  \citenamefont {Lu}, \citenamefont {Yada},\ and\ \citenamefont
  {Tanaka}}]{2024JPSCMaeda}%
  \BibitemOpen
  \bibfield  {author} {\bibinfo {author} {\bibfnamefont {K.}~\bibnamefont
  {Maeda}}, \bibinfo {author} {\bibfnamefont {B.}~\bibnamefont {Lu}}, \bibinfo
  {author} {\bibfnamefont {K.}~\bibnamefont {Yada}},\ and\ \bibinfo {author}
  {\bibfnamefont {Y.}~\bibnamefont {Tanaka}},\ }\bibfield  {title} {\bibinfo
  {title} {Theory of tunneling spectroscopy in unconventional $p$-wave
  magnet-superconductor hybrid structures},\ }\href
  {https://doi.org/10.7566/JPSJ.93.114703} {\bibfield  {journal} {\bibinfo
  {journal} {J. Phys. Soc. Jpn.}\ }\textbf {\bibinfo {volume} {93}},\ \bibinfo
  {pages} {114703} (\bibinfo {year} {2024})}\BibitemShut {NoStop}%
\bibitem [{\citenamefont {Ezawa}(2024)}]{2024PRBEzawa}%
  \BibitemOpen
  \bibfield  {author} {\bibinfo {author} {\bibfnamefont {M.}~\bibnamefont
  {Ezawa}},\ }\bibfield  {title} {\bibinfo {title} {Topological insulators and
  superconductors based on $p$-wave magnets: Electrical control and detection
  of a domain wall},\ }\href {https://doi.org/10.1103/PhysRevB.110.165429}
  {\bibfield  {journal} {\bibinfo  {journal} {Phys. Rev. B}\ }\textbf {\bibinfo
  {volume} {110}},\ \bibinfo {pages} {165429} (\bibinfo {year}
  {2024})}\BibitemShut {NoStop}%
\bibitem [{\citenamefont {Kudasov}(2024)}]{2024PRBKudasov}%
  \BibitemOpen
  \bibfield  {author} {\bibinfo {author} {\bibfnamefont {Y.~B.}\ \bibnamefont
  {Kudasov}},\ }\bibfield  {title} {\bibinfo {title} {Topological band
  structure due to modified kramers degeneracy for electrons in a helical
  magnetic field},\ }\href {https://doi.org/10.1103/PhysRevB.109.L140402}
  {\bibfield  {journal} {\bibinfo  {journal} {Phys. Rev. B}\ }\textbf {\bibinfo
  {volume} {109}},\ \bibinfo {pages} {L140402} (\bibinfo {year}
  {2024})}\BibitemShut {NoStop}%
\bibitem [{\citenamefont {Luo}\ \emph {et~al.}(2025)\citenamefont {Luo},
  \citenamefont {Hu}, \citenamefont {Hu},\ and\ \citenamefont
  {Law}}]{2025arXivLuo}%
  \BibitemOpen
  \bibfield  {author} {\bibinfo {author} {\bibfnamefont {X.-J.}\ \bibnamefont
  {Luo}}, \bibinfo {author} {\bibfnamefont {J.-X.}\ \bibnamefont {Hu}},
  \bibinfo {author} {\bibfnamefont {M.-L.}\ \bibnamefont {Hu}},\ and\ \bibinfo
  {author} {\bibfnamefont {K.}~\bibnamefont {Law}},\ }\bibfield  {title}
  {\bibinfo {title} {Spin group symmetry criteria for odd-parity magnets},\
  }\href {https://arxiv.org/abs/2510.05512} {\bibfield  {journal} {\bibinfo
  {journal} {arXiv:2510.05512}\ } (\bibinfo {year} {2025})}\BibitemShut
  {NoStop}%
\bibitem [{\citenamefont {Chakraborty}\ \emph {et~al.}(2025)\citenamefont
  {Chakraborty}, \citenamefont {Birk~Hellenes}, \citenamefont
  {Jaeschke-Ubiergo}, \citenamefont {Jungwirth}, \citenamefont
  {{\v{S}}mejkal},\ and\ \citenamefont {Sinova}}]{2025NCChakraborty}%
  \BibitemOpen
  \bibfield  {author} {\bibinfo {author} {\bibfnamefont {A.}~\bibnamefont
  {Chakraborty}}, \bibinfo {author} {\bibfnamefont {A.}~\bibnamefont
  {Birk~Hellenes}}, \bibinfo {author} {\bibfnamefont {R.}~\bibnamefont
  {Jaeschke-Ubiergo}}, \bibinfo {author} {\bibfnamefont {T.}~\bibnamefont
  {Jungwirth}}, \bibinfo {author} {\bibfnamefont {L.}~\bibnamefont
  {{\v{S}}mejkal}},\ and\ \bibinfo {author} {\bibfnamefont {J.}~\bibnamefont
  {Sinova}},\ }\bibfield  {title} {\bibinfo {title} {Highly efficient
  non-relativistic {Edelstein} effect in nodal $p$-wave magnets},\ }\href
  {https://doi.org/10.1038/s41467-025-52948-7} {\bibfield  {journal} {\bibinfo
  {journal} {Nat. Commun.}\ }\textbf {\bibinfo {volume} {16}},\ \bibinfo
  {pages} {7270} (\bibinfo {year} {2025})}\BibitemShut {NoStop}%
\bibitem [{\citenamefont {Yu}\ \emph {et~al.}(2025)\citenamefont {Yu},
  \citenamefont {Lyngby}, \citenamefont {Shishidou}, \citenamefont {Roig},
  \citenamefont {Kreisel}, \citenamefont {Weinert}, \citenamefont {Andersen},\
  and\ \citenamefont {Agterberg}}]{2025PRLYuyue}%
  \BibitemOpen
  \bibfield  {author} {\bibinfo {author} {\bibfnamefont {Y.}~\bibnamefont
  {Yu}}, \bibinfo {author} {\bibfnamefont {M.~B.}\ \bibnamefont {Lyngby}},
  \bibinfo {author} {\bibfnamefont {T.}~\bibnamefont {Shishidou}}, \bibinfo
  {author} {\bibfnamefont {M.}~\bibnamefont {Roig}}, \bibinfo {author}
  {\bibfnamefont {A.}~\bibnamefont {Kreisel}}, \bibinfo {author} {\bibfnamefont
  {M.}~\bibnamefont {Weinert}}, \bibinfo {author} {\bibfnamefont {B.~M.}\
  \bibnamefont {Andersen}},\ and\ \bibinfo {author} {\bibfnamefont {D.~F.}\
  \bibnamefont {Agterberg}},\ }\bibfield  {title} {\bibinfo {title} {Odd-parity
  magnetism driven by antiferromagnetic exchange},\ }\href
  {https://doi.org/10.1103/PhysRevLett.135.046701} {\bibfield  {journal}
  {\bibinfo  {journal} {Phys. Rev. Lett.}\ }\textbf {\bibinfo {volume} {135}},\
  \bibinfo {pages} {046701} (\bibinfo {year} {2025})}\BibitemShut {NoStop}%
\bibitem [{\citenamefont {Ezawa}(2025{\natexlab{a}})}]{2025PRBEzawa}%
  \BibitemOpen
  \bibfield  {author} {\bibinfo {author} {\bibfnamefont {M.}~\bibnamefont
  {Ezawa}},\ }\bibfield  {title} {\bibinfo {title} {Third-order and fifth-order
  nonlinear spin-current generation in $g$-wave and $i$-wave altermagnets and
  perfectly nonreciprocal spin current in $f$-wave magnets},\ }\href
  {https://doi.org/10.1103/PhysRevB.111.125420} {\bibfield  {journal} {\bibinfo
   {journal} {Phys. Rev. B}\ }\textbf {\bibinfo {volume} {111}},\ \bibinfo
  {pages} {125420} (\bibinfo {year} {2025}{\natexlab{a}})}\BibitemShut
  {NoStop}%
\bibitem [{\citenamefont {Ezawa}(2025{\natexlab{b}})}]{2025PRBEzawa2}%
  \BibitemOpen
  \bibfield  {author} {\bibinfo {author} {\bibfnamefont {M.}~\bibnamefont
  {Ezawa}},\ }\bibfield  {title} {\bibinfo {title} {Purely electrical detection
  of the spin-splitting vector in $p$-wave magnets based on linear and
  nonlinear conductivities},\ }\href
  {https://doi.org/10.1103/PhysRevB.112.125412} {\bibfield  {journal} {\bibinfo
   {journal} {Phys. Rev. B}\ }\textbf {\bibinfo {volume} {112}},\ \bibinfo
  {pages} {125412} (\bibinfo {year} {2025}{\natexlab{b}})}\BibitemShut
  {NoStop}%
\bibitem [{\citenamefont {Sun}\ \emph {et~al.}(2025)\citenamefont {Sun},
  \citenamefont {Feng}, \citenamefont {Xie}, \citenamefont {Zhou},
  \citenamefont {Hu},\ and\ \citenamefont {Law}}]{2025PRBSun}%
  \BibitemOpen
  \bibfield  {author} {\bibinfo {author} {\bibfnamefont {Z.-T.}\ \bibnamefont
  {Sun}}, \bibinfo {author} {\bibfnamefont {X.}~\bibnamefont {Feng}}, \bibinfo
  {author} {\bibfnamefont {Y.-M.}\ \bibnamefont {Xie}}, \bibinfo {author}
  {\bibfnamefont {B.~T.}\ \bibnamefont {Zhou}}, \bibinfo {author}
  {\bibfnamefont {J.-X.}\ \bibnamefont {Hu}},\ and\ \bibinfo {author}
  {\bibfnamefont {K.~T.}\ \bibnamefont {Law}},\ }\bibfield  {title} {\bibinfo
  {title} {Pseudo-ising superconductivity induced by $p$-wave magnetism},\
  }\href {https://doi.org/10.1103/PhysRevB.112.214504} {\bibfield  {journal}
  {\bibinfo  {journal} {Phys. Rev. B}\ }\textbf {\bibinfo {volume} {112}},\
  \bibinfo {pages} {214504} (\bibinfo {year} {2025})}\BibitemShut {NoStop}%
\bibitem [{\citenamefont {Jungwirth}\ \emph {et~al.}(2025)\citenamefont
  {Jungwirth}, \citenamefont {Fernandes}, \citenamefont {Fradkin},
  \citenamefont {MacDonald}, \citenamefont {Sinova},\ and\ \citenamefont
  {Šmejkal}}]{2025NewtonJungwirth}%
  \BibitemOpen
  \bibfield  {author} {\bibinfo {author} {\bibfnamefont {T.}~\bibnamefont
  {Jungwirth}}, \bibinfo {author} {\bibfnamefont {R.~M.}\ \bibnamefont
  {Fernandes}}, \bibinfo {author} {\bibfnamefont {E.}~\bibnamefont {Fradkin}},
  \bibinfo {author} {\bibfnamefont {A.~H.}\ \bibnamefont {MacDonald}}, \bibinfo
  {author} {\bibfnamefont {J.}~\bibnamefont {Sinova}},\ and\ \bibinfo {author}
  {\bibfnamefont {L.}~\bibnamefont {Šmejkal}},\ }\bibfield  {title} {\bibinfo
  {title} {Altermagnetism: An unconventional spin-ordered phase of matter},\
  }\href {https://doi.org/10.1016/j.newton.2025.100162} {\bibfield  {journal}
  {\bibinfo  {journal} {Newton}\ }\textbf {\bibinfo {volume} {1}},\ \bibinfo
  {pages} {100162} (\bibinfo {year} {2025})}\BibitemShut {NoStop}%
\bibitem [{\citenamefont {Pari}\ \emph {et~al.}(2025)\citenamefont {Pari},
  \citenamefont {Jaeschke-Ubiergo}, \citenamefont {Chakraborty}, \citenamefont
  {\ifmmode~\check{S}\else \v{S}\fi{}mejkal},\ and\ \citenamefont
  {Sinova}}]{2025PRBPari}%
  \BibitemOpen
  \bibfield  {author} {\bibinfo {author} {\bibfnamefont {N.~A.~A.}\
  \bibnamefont {Pari}}, \bibinfo {author} {\bibfnamefont {R.}~\bibnamefont
  {Jaeschke-Ubiergo}}, \bibinfo {author} {\bibfnamefont {A.}~\bibnamefont
  {Chakraborty}}, \bibinfo {author} {\bibfnamefont {L.}~\bibnamefont
  {\ifmmode~\check{S}\else \v{S}\fi{}mejkal}},\ and\ \bibinfo {author}
  {\bibfnamefont {J.}~\bibnamefont {Sinova}},\ }\bibfield  {title} {\bibinfo
  {title} {Nonrelativistic linear {Edelstein} effect in helical
  {${\mathrm{EuIn}}_{2}{\mathrm{As}}_{2}$}},\ }\href
  {https://doi.org/10.1103/PhysRevB.112.024404} {\bibfield  {journal} {\bibinfo
   {journal} {Phys. Rev. B}\ }\textbf {\bibinfo {volume} {112}},\ \bibinfo
  {pages} {024404} (\bibinfo {year} {2025})}\BibitemShut {NoStop}%
\bibitem [{\citenamefont {{Okumura}}\ \emph {et~al.}(2025)\citenamefont
  {{Okumura}}, \citenamefont {{Hirschmann}},\ and\ \citenamefont
  {{Motome}}}]{2025arXivOkumura}%
  \BibitemOpen
  \bibfield  {author} {\bibinfo {author} {\bibfnamefont {S.}~\bibnamefont
  {{Okumura}}}, \bibinfo {author} {\bibfnamefont {M.~M.}\ \bibnamefont
  {{Hirschmann}}},\ and\ \bibinfo {author} {\bibfnamefont {Y.}~\bibnamefont
  {{Motome}}},\ }\bibfield  {title} {\bibinfo {title} {Spiral-induced anomalous
  {Hall} effect from odd-parity spin-nodal lines},\ }\href
  {https://arxiv.org/abs/2512.14071} {\bibfield  {journal} {\bibinfo  {journal}
  {arXiv:2512.14071}\ } (\bibinfo {year} {2025})}\BibitemShut {NoStop}%
\bibitem [{\citenamefont {{Lin}}\ and\ \citenamefont
  {{Vila}}(2025)}]{2025arXivLin}%
  \BibitemOpen
  \bibfield  {author} {\bibinfo {author} {\bibfnamefont {Y.-P.}\ \bibnamefont
  {{Lin}}}\ and\ \bibinfo {author} {\bibfnamefont {M.}~\bibnamefont {{Vila}}},\
  }\bibfield  {title} {\bibinfo {title} {Odd-parity altermagnetism through
  sublattice currents: From {Haldane-Hubbard} model to general bipartite
  lattices},\ }\href {https://arxiv.org/abs/2503.09602} {\bibfield  {journal}
  {\bibinfo  {journal} {arXiv:2503.09602}\ } (\bibinfo {year}
  {2025})}\BibitemShut {NoStop}%
\bibitem [{\citenamefont {{Zhou}}\ \emph {et~al.}(2025)\citenamefont {{Zhou}},
  \citenamefont {{Wang}}, \citenamefont {{Ma}}, \citenamefont {{Ma}},
  \citenamefont {{Li}}, \citenamefont {{Tao}}, \citenamefont {{Zheng}},
  \citenamefont {{Yan}}, \citenamefont {{Peng}}, \citenamefont {{Shao}},
  \citenamefont {{Liu}},\ and\ \citenamefont {{Li}}}]{2025arXivZhou}%
  \BibitemOpen
  \bibfield  {author} {\bibinfo {author} {\bibfnamefont {H.}~\bibnamefont
  {{Zhou}}}, \bibinfo {author} {\bibfnamefont {M.}~\bibnamefont {{Wang}}},
  \bibinfo {author} {\bibfnamefont {Y.}~\bibnamefont {{Ma}}}, \bibinfo {author}
  {\bibfnamefont {X.}~\bibnamefont {{Ma}}}, \bibinfo {author} {\bibfnamefont
  {G.}~\bibnamefont {{Li}}}, \bibinfo {author} {\bibfnamefont {Z.}~\bibnamefont
  {{Tao}}}, \bibinfo {author} {\bibfnamefont {X.}~\bibnamefont {{Zheng}}},
  \bibinfo {author} {\bibfnamefont {L.}~\bibnamefont {{Yan}}}, \bibinfo
  {author} {\bibfnamefont {Y.}~\bibnamefont {{Peng}}}, \bibinfo {author}
  {\bibfnamefont {D.-F.}\ \bibnamefont {{Shao}}}, \bibinfo {author}
  {\bibfnamefont {B.}~\bibnamefont {{Liu}}},\ and\ \bibinfo {author}
  {\bibfnamefont {S.}~\bibnamefont {{Li}}},\ }\bibfield  {title} {\bibinfo
  {title} {Sub-spin-flop switching of a fully compensated antiferromagnet by
  magnetic field},\ }\href {https://arxiv.org/abs/2509.07351} {\bibfield
  {journal} {\bibinfo  {journal} {arXiv:2509.07351}\ } (\bibinfo {year}
  {2025})}\BibitemShut {NoStop}%
\bibitem [{\citenamefont {Song}\ \emph {et~al.}(2026)\citenamefont {Song},
  \citenamefont {Qi}, \citenamefont {Fang}, \citenamefont {Fang},\ and\
  \citenamefont {Weng}}]{2026PRXSong}%
  \BibitemOpen
  \bibfield  {author} {\bibinfo {author} {\bibfnamefont {Z.}~\bibnamefont
  {Song}}, \bibinfo {author} {\bibfnamefont {Z.}~\bibnamefont {Qi}}, \bibinfo
  {author} {\bibfnamefont {C.}~\bibnamefont {Fang}}, \bibinfo {author}
  {\bibfnamefont {Z.}~\bibnamefont {Fang}},\ and\ \bibinfo {author}
  {\bibfnamefont {H.}~\bibnamefont {Weng}},\ }\bibfield  {title} {\bibinfo
  {title} {Unified symmetry classification of magnetic orders via spin space
  groups: Prediction of coplanar even-wave phases},\ }\href
  {https://doi.org/10.1103/zy7s-j86r} {\bibfield  {journal} {\bibinfo
  {journal} {Phys. Rev. X}\ }\textbf {\bibinfo {volume} {16}},\ \bibinfo
  {pages} {031038} (\bibinfo {year} {2026})}\BibitemShut {NoStop}%
\bibitem [{\citenamefont {Zeng}\ \emph {et~al.}(2026)\citenamefont {Zeng},
  \citenamefont {Qin}, \citenamefont {Qin}, \citenamefont {Feng}, \citenamefont
  {Wu}, \citenamefont {Xu},\ and\ \citenamefont {Wang}}]{2026PRBZeng}%
  \BibitemOpen
  \bibfield  {author} {\bibinfo {author} {\bibfnamefont {M.}~\bibnamefont
  {Zeng}}, \bibinfo {author} {\bibfnamefont {Z.}~\bibnamefont {Qin}}, \bibinfo
  {author} {\bibfnamefont {L.}~\bibnamefont {Qin}}, \bibinfo {author}
  {\bibfnamefont {S.}~\bibnamefont {Feng}}, \bibinfo {author} {\bibfnamefont
  {L.}~\bibnamefont {Wu}}, \bibinfo {author} {\bibfnamefont {D.-H.}\
  \bibnamefont {Xu}},\ and\ \bibinfo {author} {\bibfnamefont {R.}~\bibnamefont
  {Wang}},\ }\bibfield  {title} {\bibinfo {title} {Odd-parity altermagnetism: A
  spin group study},\ }\href {https://doi.org/10.1103/PhysRevB.113.L220412}
  {\bibfield  {journal} {\bibinfo  {journal} {Phys. Rev. B}\ }\textbf {\bibinfo
  {volume} {113}},\ \bibinfo {pages} {L220412} (\bibinfo {year}
  {2026})}\BibitemShut {NoStop}%
\bibitem [{\citenamefont {Huang}\ \emph {et~al.}(2026)\citenamefont {Huang},
  \citenamefont {Qin}, \citenamefont {Zhan}, \citenamefont {Xu}, \citenamefont
  {Ma},\ and\ \citenamefont {Wang}}]{2026PRLHuang}%
  \BibitemOpen
  \bibfield  {author} {\bibinfo {author} {\bibfnamefont {S.}~\bibnamefont
  {Huang}}, \bibinfo {author} {\bibfnamefont {Z.}~\bibnamefont {Qin}}, \bibinfo
  {author} {\bibfnamefont {F.}~\bibnamefont {Zhan}}, \bibinfo {author}
  {\bibfnamefont {D.-H.}\ \bibnamefont {Xu}}, \bibinfo {author} {\bibfnamefont
  {D.-S.}\ \bibnamefont {Ma}},\ and\ \bibinfo {author} {\bibfnamefont
  {R.}~\bibnamefont {Wang}},\ }\bibfield  {title} {\bibinfo {title}
  {Light-induced odd-parity magnetism in conventional antiferromagnetism},\
  }\href {https://doi.org/10.1103/PhysRevLett.136.126703} {\bibfield  {journal}
  {\bibinfo  {journal} {Phys. Rev. Lett.}\ }\textbf {\bibinfo {volume} {136}},\
  \bibinfo {pages} {126703} (\bibinfo {year} {2026})}\BibitemShut {NoStop}%
\bibitem [{\citenamefont {Zhu}\ \emph {et~al.}(2026)\citenamefont {Zhu},
  \citenamefont {Zhou}, \citenamefont {Wang}, \citenamefont {Wei},\ and\
  \citenamefont {Ruan}}]{2026PRLZhu}%
  \BibitemOpen
  \bibfield  {author} {\bibinfo {author} {\bibfnamefont {T.}~\bibnamefont
  {Zhu}}, \bibinfo {author} {\bibfnamefont {D.}~\bibnamefont {Zhou}}, \bibinfo
  {author} {\bibfnamefont {H.}~\bibnamefont {Wang}}, \bibinfo {author}
  {\bibfnamefont {S.-H.}\ \bibnamefont {Wei}},\ and\ \bibinfo {author}
  {\bibfnamefont {J.}~\bibnamefont {Ruan}},\ }\bibfield  {title} {\bibinfo
  {title} {Floquet odd-parity collinear magnets},\ }\href
  {https://doi.org/10.1103/PhysRevLett.136.126704} {\bibfield  {journal}
  {\bibinfo  {journal} {Phys. Rev. Lett.}\ }\textbf {\bibinfo {volume} {136}},\
  \bibinfo {pages} {126704} (\bibinfo {year} {2026})}\BibitemShut {NoStop}%
\bibitem [{\citenamefont {Li}\ \emph {et~al.}(2026)\citenamefont {Li},
  \citenamefont {Shao},\ and\ \citenamefont {Kovalev}}]{2026PRLLi}%
  \BibitemOpen
  \bibfield  {author} {\bibinfo {author} {\bibfnamefont {B.}~\bibnamefont
  {Li}}, \bibinfo {author} {\bibfnamefont {D.-F.}\ \bibnamefont {Shao}},\ and\
  \bibinfo {author} {\bibfnamefont {A.~A.}\ \bibnamefont {Kovalev}},\
  }\bibfield  {title} {\bibinfo {title} {Floquet spin splitting and spin
  generation in antiferromagnets},\ }\href
  {https://doi.org/10.1103/PhysRevLett.136.166701} {\bibfield  {journal}
  {\bibinfo  {journal} {Phys. Rev. Lett.}\ }\textbf {\bibinfo {volume} {136}},\
  \bibinfo {pages} {166701} (\bibinfo {year} {2026})}\BibitemShut {NoStop}%
\bibitem [{\citenamefont {Liu}\ \emph {et~al.}(2026)\citenamefont {Liu},
  \citenamefont {Zhuang}, \citenamefont {Zhu}, \citenamefont {Wu},\ and\
  \citenamefont {Yan}}]{2026PRBliu}%
  \BibitemOpen
  \bibfield  {author} {\bibinfo {author} {\bibfnamefont {D.}~\bibnamefont
  {Liu}}, \bibinfo {author} {\bibfnamefont {Z.-Y.}\ \bibnamefont {Zhuang}},
  \bibinfo {author} {\bibfnamefont {D.}~\bibnamefont {Zhu}}, \bibinfo {author}
  {\bibfnamefont {Z.}~\bibnamefont {Wu}},\ and\ \bibinfo {author}
  {\bibfnamefont {Z.}~\bibnamefont {Yan}},\ }\bibfield  {title} {\bibinfo
  {title} {Light-induced odd-parity altermagnets on dimerized lattices},\
  }\href {https://doi.org/10.1103/wnqs-3djt} {\bibfield  {journal} {\bibinfo
  {journal} {Phys. Rev. B}\ }\textbf {\bibinfo {volume} {113}},\ \bibinfo
  {pages} {L060409} (\bibinfo {year} {2026})}\BibitemShut {NoStop}%
\bibitem [{\citenamefont {Fu}\ \emph {et~al.}(2026)\citenamefont {Fu},
  \citenamefont {Mondal}, \citenamefont {Liu}, \citenamefont {Tanaka},\ and\
  \citenamefont {Cayao}}]{2026PRLFuPH}%
  \BibitemOpen
  \bibfield  {author} {\bibinfo {author} {\bibfnamefont {P.-H.}\ \bibnamefont
  {Fu}}, \bibinfo {author} {\bibfnamefont {S.}~\bibnamefont {Mondal}}, \bibinfo
  {author} {\bibfnamefont {J.-F.}\ \bibnamefont {Liu}}, \bibinfo {author}
  {\bibfnamefont {Y.}~\bibnamefont {Tanaka}},\ and\ \bibinfo {author}
  {\bibfnamefont {J.}~\bibnamefont {Cayao}},\ }\bibfield  {title} {\bibinfo
  {title} {Floquet engineering spin triplet states in unconventional magnets},\
  }\href {https://doi.org/10.1103/PhysRevLett.136.066703} {\bibfield  {journal}
  {\bibinfo  {journal} {Phys. Rev. Lett.}\ }\textbf {\bibinfo {volume} {136}},\
  \bibinfo {pages} {066703} (\bibinfo {year} {2026})}\BibitemShut {NoStop}%
\bibitem [{\citenamefont {Dsouza}\ \emph {et~al.}(2026)\citenamefont {Dsouza},
  \citenamefont {Kreisel}, \citenamefont {Andersen}, \citenamefont
  {Agterberg},\ and\ \citenamefont {Christensen}}]{2026PRBDsouza}%
  \BibitemOpen
  \bibfield  {author} {\bibinfo {author} {\bibfnamefont {R.}~\bibnamefont
  {Dsouza}}, \bibinfo {author} {\bibfnamefont {A.}~\bibnamefont {Kreisel}},
  \bibinfo {author} {\bibfnamefont {B.~M.}\ \bibnamefont {Andersen}}, \bibinfo
  {author} {\bibfnamefont {D.~F.}\ \bibnamefont {Agterberg}},\ and\ \bibinfo
  {author} {\bibfnamefont {M.~H.}\ \bibnamefont {Christensen}},\ }\bibfield
  {title} {\bibinfo {title} {Odd-parity magnetism in {Fe}-based superconductors
  with coplanar magnetic order},\ }\href
  {https://doi.org/10.1103/PhysRevB.113.144509} {\bibfield  {journal} {\bibinfo
   {journal} {Phys. Rev. B}\ }\textbf {\bibinfo {volume} {113}},\ \bibinfo
  {pages} {144509} (\bibinfo {year} {2026})}\BibitemShut {NoStop}%
\bibitem [{\citenamefont {Khodas}\ \emph {et~al.}(2026)\citenamefont {Khodas},
  \citenamefont {Šmejkal},\ and\ \citenamefont {Mazin}}]{2026arXivMaxim}%
  \BibitemOpen
  \bibfield  {author} {\bibinfo {author} {\bibfnamefont {M.}~\bibnamefont
  {Khodas}}, \bibinfo {author} {\bibfnamefont {L.}~\bibnamefont {Šmejkal}},\
  and\ \bibinfo {author} {\bibfnamefont {I.~I.}\ \bibnamefont {Mazin}},\
  }\bibfield  {title} {\bibinfo {title} {Nonrelativistic-ising
  superconductivity in $p$-wave magnets},\ }\href
  {https://arxiv.org/abs/2601.19829} {\bibfield  {journal} {\bibinfo  {journal}
  {arXiv:2601.19829}\ } (\bibinfo {year} {2026})}\BibitemShut {NoStop}%
\bibitem [{\citenamefont {Luo}\ \emph {et~al.}(2026)\citenamefont {Luo},
  \citenamefont {Sun}, \citenamefont {Feng}, \citenamefont {Tian},\ and\
  \citenamefont {Law}}]{2026arXivLuo}%
  \BibitemOpen
  \bibfield  {author} {\bibinfo {author} {\bibfnamefont {X.-J.}\ \bibnamefont
  {Luo}}, \bibinfo {author} {\bibfnamefont {Z.-T.}\ \bibnamefont {Sun}},
  \bibinfo {author} {\bibfnamefont {X.}~\bibnamefont {Feng}}, \bibinfo {author}
  {\bibfnamefont {M.}~\bibnamefont {Tian}},\ and\ \bibinfo {author}
  {\bibfnamefont {K.~T.}\ \bibnamefont {Law}},\ }\bibfield  {title} {\bibinfo
  {title} {Hidden {Zeeman} field in odd-parity magnets: An ideal platform for
  topological superconductivity},\ }\href {https://arxiv.org/abs/2603.15147}
  {\bibfield  {journal} {\bibinfo  {journal} {arXiv:2603.15147}\ } (\bibinfo
  {year} {2026})}\BibitemShut {NoStop}%
\bibitem [{\citenamefont {Mitscherling}\ \emph {et~al.}(2026)\citenamefont
  {Mitscherling}, \citenamefont {Priessnitz}, \citenamefont {Geschner},\ and\
  \citenamefont {Šmejkal}}]{2026arXivJohannes}%
  \BibitemOpen
  \bibfield  {author} {\bibinfo {author} {\bibfnamefont {J.}~\bibnamefont
  {Mitscherling}}, \bibinfo {author} {\bibfnamefont {J.}~\bibnamefont
  {Priessnitz}}, \bibinfo {author} {\bibfnamefont {C.~K.}\ \bibnamefont
  {Geschner}},\ and\ \bibinfo {author} {\bibfnamefont {L.}~\bibnamefont
  {Šmejkal}},\ }\bibfield  {title} {\bibinfo {title} {Microscopic origin of
  $p$-wave magnetism},\ }\href {https://arxiv.org/abs/2603.09736} {\bibfield
  {journal} {\bibinfo  {journal} {arXiv:2603.09736}\ } (\bibinfo {year}
  {2026})}\BibitemShut {NoStop}%
\bibitem [{\citenamefont {Priessnitz}\ \emph {et~al.}(2026)\citenamefont
  {Priessnitz}, \citenamefont {Hellenes}, \citenamefont {Comin},\ and\
  \citenamefont {Šmejkal}}]{2026arXivJan}%
  \BibitemOpen
  \bibfield  {author} {\bibinfo {author} {\bibfnamefont {J.}~\bibnamefont
  {Priessnitz}}, \bibinfo {author} {\bibfnamefont {A.~B.}\ \bibnamefont
  {Hellenes}}, \bibinfo {author} {\bibfnamefont {R.}~\bibnamefont {Comin}},\
  and\ \bibinfo {author} {\bibfnamefont {L.}~\bibnamefont {Šmejkal}},\
  }\bibfield  {title} {\bibinfo {title} {Ferroelectric $p$-wave magnets},\
  }\href {https://arxiv.org/abs/2603.19107} {\bibfield  {journal} {\bibinfo
  {journal} {arXiv:2603.19107}\ } (\bibinfo {year} {2026})}\BibitemShut
  {NoStop}%
\bibitem [{\citenamefont {Neumann}\ \emph {et~al.}(2026)\citenamefont
  {Neumann}, \citenamefont {Jaeschke-Ubiergo}, \citenamefont {Zarzuela},
  \citenamefont {Šmejkal}, \citenamefont {Sinova},\ and\ \citenamefont
  {Mook}}]{2026arXivRobin}%
  \BibitemOpen
  \bibfield  {author} {\bibinfo {author} {\bibfnamefont {R.~R.}\ \bibnamefont
  {Neumann}}, \bibinfo {author} {\bibfnamefont {R.}~\bibnamefont
  {Jaeschke-Ubiergo}}, \bibinfo {author} {\bibfnamefont {R.}~\bibnamefont
  {Zarzuela}}, \bibinfo {author} {\bibfnamefont {L.}~\bibnamefont {Šmejkal}},
  \bibinfo {author} {\bibfnamefont {J.}~\bibnamefont {Sinova}},\ and\ \bibinfo
  {author} {\bibfnamefont {A.}~\bibnamefont {Mook}},\ }\bibfield  {title}
  {\bibinfo {title} {Odd-parity-wave magnons and nonrelativistic thermal
  {Edelstein} effect},\ }\href {https://arxiv.org/abs/2603.05415} {\bibfield
  {journal} {\bibinfo  {journal} {arXiv:2603.05415}\ } (\bibinfo {year}
  {2026})}\BibitemShut {NoStop}%
\bibitem [{\citenamefont {Zhang}\ \emph {et~al.}(2026)\citenamefont {Zhang},
  \citenamefont {Li}, \citenamefont {Cheng}, \citenamefont {Fan}, \citenamefont
  {Yin}, \citenamefont {Gao}, \citenamefont {Liu}, \citenamefont {Cui},
  \citenamefont {Yin}, \citenamefont {Zhao}, \citenamefont {Lin}, \citenamefont
  {Liu}, \citenamefont {Ye}, \citenamefont {Huang}, \citenamefont {Qiao},
  \citenamefont {Xie}, \citenamefont {Miao}, \citenamefont {Wu}, \citenamefont
  {Liu}, \citenamefont {Cao},\ and\ \citenamefont {Chen}}]{2026arXivZhang}%
  \BibitemOpen
  \bibfield  {author} {\bibinfo {author} {\bibfnamefont {F.}~\bibnamefont
  {Zhang}}, \bibinfo {author} {\bibfnamefont {H.}~\bibnamefont {Li}}, \bibinfo
  {author} {\bibfnamefont {X.}~\bibnamefont {Cheng}}, \bibinfo {author}
  {\bibfnamefont {Y.}~\bibnamefont {Fan}}, \bibinfo {author} {\bibfnamefont
  {Y.}~\bibnamefont {Yin}}, \bibinfo {author} {\bibfnamefont {Y.}~\bibnamefont
  {Gao}}, \bibinfo {author} {\bibfnamefont {Z.}~\bibnamefont {Liu}}, \bibinfo
  {author} {\bibfnamefont {S.}~\bibnamefont {Cui}}, \bibinfo {author}
  {\bibfnamefont {Z.}~\bibnamefont {Yin}}, \bibinfo {author} {\bibfnamefont
  {Y.}~\bibnamefont {Zhao}}, \bibinfo {author} {\bibfnamefont {J.}~\bibnamefont
  {Lin}}, \bibinfo {author} {\bibfnamefont {Z.}~\bibnamefont {Liu}}, \bibinfo
  {author} {\bibfnamefont {M.}~\bibnamefont {Ye}}, \bibinfo {author}
  {\bibfnamefont {Y.}~\bibnamefont {Huang}}, \bibinfo {author} {\bibfnamefont
  {S.}~\bibnamefont {Qiao}}, \bibinfo {author} {\bibfnamefont {W.}~\bibnamefont
  {Xie}}, \bibinfo {author} {\bibfnamefont {P.}~\bibnamefont {Miao}}, \bibinfo
  {author} {\bibfnamefont {H.}~\bibnamefont {Wu}}, \bibinfo {author}
  {\bibfnamefont {J.}~\bibnamefont {Liu}}, \bibinfo {author} {\bibfnamefont
  {G.}~\bibnamefont {Cao}},\ and\ \bibinfo {author} {\bibfnamefont
  {C.}~\bibnamefont {Chen}},\ }\bibfield  {title} {\bibinfo {title} {Odd spin
  symmetry and anisotropy switching in $p$-wave magnet {CeNiAsO}},\ }\href
  {https://arxiv.org/abs/2605.28701} {\bibfield  {journal} {\bibinfo  {journal}
  {arXiv:2605.28701}\ } (\bibinfo {year} {2026})}\BibitemShut {NoStop}%
\bibitem [{\citenamefont {Gallego}\ \emph {et~al.}(2016)\citenamefont
  {Gallego}, \citenamefont {Perez-Mato}, \citenamefont {Elcoro}, \citenamefont
  {Tasci}, \citenamefont {Hanson}, \citenamefont {Momma}, \citenamefont
  {Aroyo},\ and\ \citenamefont {Madariaga}}]{2016ACGallego}%
  \BibitemOpen
  \bibfield  {author} {\bibinfo {author} {\bibfnamefont {S.~V.}\ \bibnamefont
  {Gallego}}, \bibinfo {author} {\bibfnamefont {J.~M.}\ \bibnamefont
  {Perez-Mato}}, \bibinfo {author} {\bibfnamefont {L.}~\bibnamefont {Elcoro}},
  \bibinfo {author} {\bibfnamefont {E.~S.}\ \bibnamefont {Tasci}}, \bibinfo
  {author} {\bibfnamefont {R.~M.}\ \bibnamefont {Hanson}}, \bibinfo {author}
  {\bibfnamefont {K.}~\bibnamefont {Momma}}, \bibinfo {author} {\bibfnamefont
  {M.~I.}\ \bibnamefont {Aroyo}},\ and\ \bibinfo {author} {\bibfnamefont
  {G.}~\bibnamefont {Madariaga}},\ }\bibfield  {title} {\bibinfo {title}
  {{MAGNDATA}: towards a database of magnetic structures. {I.} the commensurate
  case},\ }\href {https://doi.org/10.1107/S1600576716012863} {\bibfield
  {journal} {\bibinfo  {journal} {J. Appl. Crystallogr.}\ }\textbf {\bibinfo
  {volume} {49}},\ \bibinfo {pages} {1750} (\bibinfo {year}
  {2016})}\BibitemShut {NoStop}%
\bibitem [{Sup()}]{SuppMater}%
  \BibitemOpen
  \href@noop {} {}\bibinfo {note} {See Supplemental Material at
  http://link.aps.org/xxx, which includes a detailed description of model
  hamiltonian, computational methods, material candidates, supplementary tables
  and figures, and
  Refs.~\cite{zhang2022magnetictb,VASP,PAWa,PAWb,GGA,PBE,DFT-D3,Wannier90,2004PRLYao,2007PRBYates,1992PRLSuzulki,1995PRBGuo,1999PRBRavindran,2016ACGallego,2026arXivYu,1973AIPCPEib}}\BibitemShut
  {NoStop}%
\bibitem [{\citenamefont {Zhou}\ \emph {et~al.}(2023)\citenamefont {Zhou},
  \citenamefont {Feng}, \citenamefont {Li},\ and\ \citenamefont
  {Yao}}]{2023NLZhou}%
  \BibitemOpen
  \bibfield  {author} {\bibinfo {author} {\bibfnamefont {X.}~\bibnamefont
  {Zhou}}, \bibinfo {author} {\bibfnamefont {W.}~\bibnamefont {Feng}}, \bibinfo
  {author} {\bibfnamefont {Y.}~\bibnamefont {Li}},\ and\ \bibinfo {author}
  {\bibfnamefont {Y.}~\bibnamefont {Yao}},\ }\bibfield  {title} {\bibinfo
  {title} {Spin-chirality-driven quantum anomalous and quantum topological
  {Hall} effects in chiral magnets},\ }\href
  {https://doi.org/10.1021/acs.nanolett.3c01413} {\bibfield  {journal}
  {\bibinfo  {journal} {Nano Lett.}\ }\textbf {\bibinfo {volume} {23}},\
  \bibinfo {pages} {10628} (\bibinfo {year} {2023})}\BibitemShut {NoStop}%
\bibitem [{\citenamefont {Nayak}\ \emph {et~al.}(2016)\citenamefont {Nayak},
  \citenamefont {Fischer}, \citenamefont {Sun}, \citenamefont {Yan},
  \citenamefont {Karel}, \citenamefont {Komarek}, \citenamefont {Shekhar},
  \citenamefont {Kumar}, \citenamefont {Schnelle}, \citenamefont {K{\"u}bler},
  \citenamefont {Felser},\ and\ \citenamefont {Parkin}}]{2016SANayak}%
  \BibitemOpen
  \bibfield  {author} {\bibinfo {author} {\bibfnamefont {A.~K.}\ \bibnamefont
  {Nayak}}, \bibinfo {author} {\bibfnamefont {J.~E.}\ \bibnamefont {Fischer}},
  \bibinfo {author} {\bibfnamefont {Y.}~\bibnamefont {Sun}}, \bibinfo {author}
  {\bibfnamefont {B.}~\bibnamefont {Yan}}, \bibinfo {author} {\bibfnamefont
  {J.}~\bibnamefont {Karel}}, \bibinfo {author} {\bibfnamefont {A.~C.}\
  \bibnamefont {Komarek}}, \bibinfo {author} {\bibfnamefont {C.}~\bibnamefont
  {Shekhar}}, \bibinfo {author} {\bibfnamefont {N.}~\bibnamefont {Kumar}},
  \bibinfo {author} {\bibfnamefont {W.}~\bibnamefont {Schnelle}}, \bibinfo
  {author} {\bibfnamefont {J.}~\bibnamefont {K{\"u}bler}}, \bibinfo {author}
  {\bibfnamefont {C.}~\bibnamefont {Felser}},\ and\ \bibinfo {author}
  {\bibfnamefont {S.~S.~P.}\ \bibnamefont {Parkin}},\ }\bibfield  {title}
  {\bibinfo {title} {Large anomalous {Hall} effect driven by a nonvanishing
  {Berry} curvature in the noncolinear antiferromagnet {Mn$_3$Ge}},\ }\href
  {https://doi.org/10.1126/sciadv.1501870} {\bibfield  {journal} {\bibinfo
  {journal} {Sci. Adv.}\ }\textbf {\bibinfo {volume} {2}},\ \bibinfo {pages}
  {e1501870} (\bibinfo {year} {2016})}\BibitemShut {NoStop}%
\bibitem [{\citenamefont {Takeuchi}\ \emph {et~al.}(2021)\citenamefont
  {Takeuchi}, \citenamefont {Yamane}, \citenamefont {Yoon}, \citenamefont
  {Itoh}, \citenamefont {Jinnai}, \citenamefont {Kanai}, \citenamefont {Ieda},
  \citenamefont {Fukami},\ and\ \citenamefont {Ohno}}]{2021NMTakeuchi}%
  \BibitemOpen
  \bibfield  {author} {\bibinfo {author} {\bibfnamefont {Y.}~\bibnamefont
  {Takeuchi}}, \bibinfo {author} {\bibfnamefont {Y.}~\bibnamefont {Yamane}},
  \bibinfo {author} {\bibfnamefont {J.-Y.}\ \bibnamefont {Yoon}}, \bibinfo
  {author} {\bibfnamefont {R.}~\bibnamefont {Itoh}}, \bibinfo {author}
  {\bibfnamefont {B.}~\bibnamefont {Jinnai}}, \bibinfo {author} {\bibfnamefont
  {S.}~\bibnamefont {Kanai}}, \bibinfo {author} {\bibfnamefont
  {J.}~\bibnamefont {Ieda}}, \bibinfo {author} {\bibfnamefont {S.}~\bibnamefont
  {Fukami}},\ and\ \bibinfo {author} {\bibfnamefont {H.}~\bibnamefont {Ohno}},\
  }\bibfield  {title} {\bibinfo {title} {Chiral-spin rotation of non-collinear
  antiferromagnet by spin--orbit torque},\ }\href
  {https://doi.org/10.1038/s41563-021-01005-3} {\bibfield  {journal} {\bibinfo
  {journal} {Nat. Mater.}\ }\textbf {\bibinfo {volume} {20}},\ \bibinfo {pages}
  {1364} (\bibinfo {year} {2021})}\BibitemShut {NoStop}%
\bibitem [{\citenamefont {Yu}\ \emph {et~al.}(2026)\citenamefont {Yu},
  \citenamefont {Chen}, \citenamefont {Zhu}, \citenamefont {Li}, \citenamefont
  {Xiong}, \citenamefont {Li}, \citenamefont {Liu},\ and\ \citenamefont
  {Liu}}]{2026arXivYu}%
  \BibitemOpen
  \bibfield  {author} {\bibinfo {author} {\bibfnamefont {Y.}~\bibnamefont
  {Yu}}, \bibinfo {author} {\bibfnamefont {X.}~\bibnamefont {Chen}}, \bibinfo
  {author} {\bibfnamefont {Y.}~\bibnamefont {Zhu}}, \bibinfo {author}
  {\bibfnamefont {Y.}~\bibnamefont {Li}}, \bibinfo {author} {\bibfnamefont
  {R.}~\bibnamefont {Xiong}}, \bibinfo {author} {\bibfnamefont
  {J.}~\bibnamefont {Li}}, \bibinfo {author} {\bibfnamefont {Y.}~\bibnamefont
  {Liu}},\ and\ \bibinfo {author} {\bibfnamefont {Q.}~\bibnamefont {Liu}},\
  }\bibfield  {title} {\bibinfo {title} {Identifying oriented spin space groups
  and related physical properties using an online platform findspingroup},\
  }\href {https://arxiv.org/abs/2604.21397} {\bibfield  {journal} {\bibinfo
  {journal} {arXiv:2604.21397}\ } (\bibinfo {year} {2026})}\BibitemShut
  {NoStop}%
\bibitem [{\citenamefont {Feng}\ \emph {et~al.}(2015)\citenamefont {Feng},
  \citenamefont {Guo}, \citenamefont {Zhou}, \citenamefont {Yao},\ and\
  \citenamefont {Niu}}]{2015PRBFeng}%
  \BibitemOpen
  \bibfield  {author} {\bibinfo {author} {\bibfnamefont {W.}~\bibnamefont
  {Feng}}, \bibinfo {author} {\bibfnamefont {G.-Y.}\ \bibnamefont {Guo}},
  \bibinfo {author} {\bibfnamefont {J.}~\bibnamefont {Zhou}}, \bibinfo {author}
  {\bibfnamefont {Y.}~\bibnamefont {Yao}},\ and\ \bibinfo {author}
  {\bibfnamefont {Q.}~\bibnamefont {Niu}},\ }\bibfield  {title} {\bibinfo
  {title} {Large magneto-optical {Kerr} effect in noncollinear antiferromagnets
  {${\mathrm{Mn}}_{3}X\phantom{\rule{0.28em}{0ex}}(X=\mathrm{Rh},\phantom{\rule{0.28em}{0ex}}\mathrm{Ir},\phantom{\rule{0.28em}{0ex}}\mathrm{Pt})$}},\
  }\href {https://doi.org/10.1103/PhysRevB.92.144426} {\bibfield  {journal}
  {\bibinfo  {journal} {Phys. Rev. B}\ }\textbf {\bibinfo {volume} {92}},\
  \bibinfo {pages} {144426} (\bibinfo {year} {2015})}\BibitemShut {NoStop}%
\bibitem [{\citenamefont {Zhou}\ \emph {et~al.}(2019)\citenamefont {Zhou},
  \citenamefont {Hanke}, \citenamefont {Feng}, \citenamefont {Li},
  \citenamefont {Guo}, \citenamefont {Yao}, \citenamefont {Bl\"ugel},\ and\
  \citenamefont {Mokrousov}}]{2019PRBZhou}%
  \BibitemOpen
  \bibfield  {author} {\bibinfo {author} {\bibfnamefont {X.}~\bibnamefont
  {Zhou}}, \bibinfo {author} {\bibfnamefont {J.-P.}\ \bibnamefont {Hanke}},
  \bibinfo {author} {\bibfnamefont {W.}~\bibnamefont {Feng}}, \bibinfo {author}
  {\bibfnamefont {F.}~\bibnamefont {Li}}, \bibinfo {author} {\bibfnamefont
  {G.-Y.}\ \bibnamefont {Guo}}, \bibinfo {author} {\bibfnamefont
  {Y.}~\bibnamefont {Yao}}, \bibinfo {author} {\bibfnamefont {S.}~\bibnamefont
  {Bl\"ugel}},\ and\ \bibinfo {author} {\bibfnamefont {Y.}~\bibnamefont
  {Mokrousov}},\ }\bibfield  {title} {\bibinfo {title} {Spin-order dependent
  anomalous {Hall} effect and magneto-optical effect in the noncollinear
  antiferromagnets {${\mathrm{Mn}}_{3}X\mathrm{N}$} with {$X$=Ga, Zn, Ag, or
  Ni}},\ }\href {https://doi.org/10.1103/PhysRevB.99.104428} {\bibfield
  {journal} {\bibinfo  {journal} {Phys. Rev. B}\ }\textbf {\bibinfo {volume}
  {99}},\ \bibinfo {pages} {104428} (\bibinfo {year} {2019})}\BibitemShut
  {NoStop}%
\bibitem [{\citenamefont {Huang}\ \emph {et~al.}(2017)\citenamefont {Huang},
  \citenamefont {Clark}, \citenamefont {Navarro-Moratalla}, \citenamefont
  {Klein}, \citenamefont {Cheng}, \citenamefont {Seyler}, \citenamefont
  {Zhong}, \citenamefont {Schmidgall}, \citenamefont {McGuire}, \citenamefont
  {Cobden}, \citenamefont {Yao}, \citenamefont {Xiao}, \citenamefont
  {Jarillo-Herrero},\ and\ \citenamefont {Xu}}]{2017NatureHuang}%
  \BibitemOpen
  \bibfield  {author} {\bibinfo {author} {\bibfnamefont {B.}~\bibnamefont
  {Huang}}, \bibinfo {author} {\bibfnamefont {G.}~\bibnamefont {Clark}},
  \bibinfo {author} {\bibfnamefont {E.}~\bibnamefont {Navarro-Moratalla}},
  \bibinfo {author} {\bibfnamefont {D.~R.}\ \bibnamefont {Klein}}, \bibinfo
  {author} {\bibfnamefont {R.}~\bibnamefont {Cheng}}, \bibinfo {author}
  {\bibfnamefont {K.~L.}\ \bibnamefont {Seyler}}, \bibinfo {author}
  {\bibfnamefont {D.}~\bibnamefont {Zhong}}, \bibinfo {author} {\bibfnamefont
  {E.}~\bibnamefont {Schmidgall}}, \bibinfo {author} {\bibfnamefont {M.~A.}\
  \bibnamefont {McGuire}}, \bibinfo {author} {\bibfnamefont {D.~H.}\
  \bibnamefont {Cobden}}, \bibinfo {author} {\bibfnamefont {W.}~\bibnamefont
  {Yao}}, \bibinfo {author} {\bibfnamefont {D.}~\bibnamefont {Xiao}}, \bibinfo
  {author} {\bibfnamefont {P.}~\bibnamefont {Jarillo-Herrero}},\ and\ \bibinfo
  {author} {\bibfnamefont {X.}~\bibnamefont {Xu}},\ }\bibfield  {title}
  {\bibinfo {title} {Layer-dependent ferromagnetism in a van der {Waals}
  crystal down to the monolayer limit},\ }\href
  {https://doi.org/10.1038/nature22391} {\bibfield  {journal} {\bibinfo
  {journal} {Nature}\ }\textbf {\bibinfo {volume} {546}},\ \bibinfo {pages}
  {270} (\bibinfo {year} {2017})}\BibitemShut {NoStop}%
\bibitem [{\citenamefont {Zhou}\ \emph {et~al.}(2017)\citenamefont {Zhou},
  \citenamefont {Feng}, \citenamefont {Li},\ and\ \citenamefont
  {Yao}}]{2017NanoSZhou}%
  \BibitemOpen
  \bibfield  {author} {\bibinfo {author} {\bibfnamefont {X.}~\bibnamefont
  {Zhou}}, \bibinfo {author} {\bibfnamefont {W.}~\bibnamefont {Feng}}, \bibinfo
  {author} {\bibfnamefont {F.}~\bibnamefont {Li}},\ and\ \bibinfo {author}
  {\bibfnamefont {Y.}~\bibnamefont {Yao}},\ }\bibfield  {title} {\bibinfo
  {title} {Large magneto-optical effects in hole-doped blue phosphorene and
  gray arsenene},\ }\href {https://doi.org/10.1039/C7NR05088G} {\bibfield
  {journal} {\bibinfo  {journal} {Nanoscale}\ }\textbf {\bibinfo {volume}
  {9}},\ \bibinfo {pages} {17405} (\bibinfo {year} {2017})}\BibitemShut
  {NoStop}%
\bibitem [{\citenamefont {Zhang}\ \emph {et~al.}(2022)\citenamefont {Zhang},
  \citenamefont {Yu}, \citenamefont {Liu},\ and\ \citenamefont
  {Yao}}]{zhang2022magnetictb}%
  \BibitemOpen
  \bibfield  {author} {\bibinfo {author} {\bibfnamefont {Z.}~\bibnamefont
  {Zhang}}, \bibinfo {author} {\bibfnamefont {Z.-M.}\ \bibnamefont {Yu}},
  \bibinfo {author} {\bibfnamefont {G.-B.}\ \bibnamefont {Liu}},\ and\ \bibinfo
  {author} {\bibfnamefont {Y.}~\bibnamefont {Yao}},\ }\bibfield  {title}
  {\bibinfo {title} {{MagneticTB}: A package for tight-binding model of
  magnetic and non-magnetic materials},\ }\href
  {https://doi.org/10.1016/j.cpc.2021.108153} {\bibfield  {journal} {\bibinfo
  {journal} {Comput. Phys. Commun.}\ }\textbf {\bibinfo {volume} {270}},\
  \bibinfo {pages} {108153} (\bibinfo {year} {2022})}\BibitemShut {NoStop}%
\bibitem [{\citenamefont {Kresse}\ and\ \citenamefont
  {Furthm\"uller}(1996)}]{VASP}%
  \BibitemOpen
  \bibfield  {author} {\bibinfo {author} {\bibfnamefont {G.}~\bibnamefont
  {Kresse}}\ and\ \bibinfo {author} {\bibfnamefont {J.}~\bibnamefont
  {Furthm\"uller}},\ }\bibfield  {title} {\bibinfo {title} {Efficient iterative
  schemes for ab initio total-energy calculations using a plane-wave basis
  set},\ }\href {https://doi.org/10.1103/PhysRevB.54.11169} {\bibfield
  {journal} {\bibinfo  {journal} {Phys. Rev. B}\ }\textbf {\bibinfo {volume}
  {54}},\ \bibinfo {pages} {11169} (\bibinfo {year} {1996})}\BibitemShut
  {NoStop}%
\bibitem [{\citenamefont {Bl\"ochl}(1994)}]{PAWa}%
  \BibitemOpen
  \bibfield  {author} {\bibinfo {author} {\bibfnamefont {P.~E.}\ \bibnamefont
  {Bl\"ochl}},\ }\bibfield  {title} {\bibinfo {title} {Projector augmented-wave
  method},\ }\href {https://doi.org/10.1103/PhysRevB.50.17953} {\bibfield
  {journal} {\bibinfo  {journal} {Phys. Rev. B}\ }\textbf {\bibinfo {volume}
  {50}},\ \bibinfo {pages} {17953} (\bibinfo {year} {1994})}\BibitemShut
  {NoStop}%
\bibitem [{\citenamefont {Kresse}\ and\ \citenamefont {Joubert}(1999)}]{PAWb}%
  \BibitemOpen
  \bibfield  {author} {\bibinfo {author} {\bibfnamefont {G.}~\bibnamefont
  {Kresse}}\ and\ \bibinfo {author} {\bibfnamefont {D.}~\bibnamefont
  {Joubert}},\ }\bibfield  {title} {\bibinfo {title} {From ultrasoft
  pseudopotentials to the projector augmented-wave method},\ }\href
  {https://doi.org/10.1103/PhysRevB.59.1758} {\bibfield  {journal} {\bibinfo
  {journal} {Phys. Rev. B}\ }\textbf {\bibinfo {volume} {59}},\ \bibinfo
  {pages} {1758} (\bibinfo {year} {1999})}\BibitemShut {NoStop}%
\bibitem [{\citenamefont {Perdew}\ and\ \citenamefont {Yue}(1986)}]{GGA}%
  \BibitemOpen
  \bibfield  {author} {\bibinfo {author} {\bibfnamefont {J.~P.}\ \bibnamefont
  {Perdew}}\ and\ \bibinfo {author} {\bibfnamefont {W.}~\bibnamefont {Yue}},\
  }\bibfield  {title} {\bibinfo {title} {Accurate and simple density functional
  for the electronic exchange energy: Generalized gradient approximation},\
  }\href {https://doi.org/10.1103/PhysRevB.33.8800} {\bibfield  {journal}
  {\bibinfo  {journal} {Phys. Rev. B}\ }\textbf {\bibinfo {volume} {33}},\
  \bibinfo {pages} {8800} (\bibinfo {year} {1986})}\BibitemShut {NoStop}%
\bibitem [{\citenamefont {Perdew}\ \emph {et~al.}(1996)\citenamefont {Perdew},
  \citenamefont {Burke},\ and\ \citenamefont {Ernzerhof}}]{PBE}%
  \BibitemOpen
  \bibfield  {author} {\bibinfo {author} {\bibfnamefont {J.~P.}\ \bibnamefont
  {Perdew}}, \bibinfo {author} {\bibfnamefont {K.}~\bibnamefont {Burke}},\ and\
  \bibinfo {author} {\bibfnamefont {M.}~\bibnamefont {Ernzerhof}},\ }\bibfield
  {title} {\bibinfo {title} {Generalized gradient approximation made simple},\
  }\href {https://doi.org/10.1103/PhysRevLett.77.3865} {\bibfield  {journal}
  {\bibinfo  {journal} {Phys. Rev. Lett.}\ }\textbf {\bibinfo {volume} {77}},\
  \bibinfo {pages} {3865} (\bibinfo {year} {1996})}\BibitemShut {NoStop}%
\bibitem [{\citenamefont {Grimme}\ \emph {et~al.}(2010)\citenamefont {Grimme},
  \citenamefont {Antony}, \citenamefont {Ehrlich},\ and\ \citenamefont
  {Krieg}}]{DFT-D3}%
  \BibitemOpen
  \bibfield  {author} {\bibinfo {author} {\bibfnamefont {S.}~\bibnamefont
  {Grimme}}, \bibinfo {author} {\bibfnamefont {J.}~\bibnamefont {Antony}},
  \bibinfo {author} {\bibfnamefont {S.}~\bibnamefont {Ehrlich}},\ and\ \bibinfo
  {author} {\bibfnamefont {H.}~\bibnamefont {Krieg}},\ }\bibfield  {title}
  {\bibinfo {title} {A consistent and accurate ab initio parametrization of
  density functional dispersion correction ({DFT-D}) for the 94 elements
  {H-Pu}},\ }\href {https://doi.org/10.1063/1.3382344} {\bibfield  {journal}
  {\bibinfo  {journal} {J. Chem. Phys.}\ }\textbf {\bibinfo {volume} {132}},\
  \bibinfo {pages} {154104} (\bibinfo {year} {2010})}\BibitemShut {NoStop}%
\bibitem [{\citenamefont {Pizzi}\ \emph {et~al.}(2020)\citenamefont {Pizzi}
  \emph {et~al.}}]{Wannier90}%
  \BibitemOpen
  \bibfield  {author} {\bibinfo {author} {\bibfnamefont {G.}~\bibnamefont
  {Pizzi}} \emph {et~al.},\ }\bibfield  {title} {\bibinfo {title} {Wannier90 as
  a community code: new features and applications},\ }\href
  {https://doi.org/10.1088/1361-648X/ab51ff} {\bibfield  {journal} {\bibinfo
  {journal} {J. Phys.: Condens. Matter}\ }\textbf {\bibinfo {volume} {32}},\
  \bibinfo {pages} {165902} (\bibinfo {year} {2020})}\BibitemShut {NoStop}%
\bibitem [{\citenamefont {Yao}\ \emph {et~al.}(2004)\citenamefont {Yao},
  \citenamefont {Kleinman}, \citenamefont {MacDonald}, \citenamefont {Sinova},
  \citenamefont {Jungwirth}, \citenamefont {Wang}, \citenamefont {Wang},\ and\
  \citenamefont {Niu}}]{2004PRLYao}%
  \BibitemOpen
  \bibfield  {author} {\bibinfo {author} {\bibfnamefont {Y.}~\bibnamefont
  {Yao}}, \bibinfo {author} {\bibfnamefont {L.}~\bibnamefont {Kleinman}},
  \bibinfo {author} {\bibfnamefont {A.~H.}\ \bibnamefont {MacDonald}}, \bibinfo
  {author} {\bibfnamefont {J.}~\bibnamefont {Sinova}}, \bibinfo {author}
  {\bibfnamefont {T.}~\bibnamefont {Jungwirth}}, \bibinfo {author}
  {\bibfnamefont {D.-s.}\ \bibnamefont {Wang}}, \bibinfo {author}
  {\bibfnamefont {E.}~\bibnamefont {Wang}},\ and\ \bibinfo {author}
  {\bibfnamefont {Q.}~\bibnamefont {Niu}},\ }\bibfield  {title} {\bibinfo
  {title} {First principles calculation of anomalous {Hall} conductivity in
  ferromagnetic bcc fe},\ }\href
  {https://doi.org/10.1103/PhysRevLett.92.037204} {\bibfield  {journal}
  {\bibinfo  {journal} {Phys. Rev. Lett.}\ }\textbf {\bibinfo {volume} {92}},\
  \bibinfo {pages} {037204} (\bibinfo {year} {2004})}\BibitemShut {NoStop}%
\bibitem [{\citenamefont {Yates}\ \emph {et~al.}(2007)\citenamefont {Yates},
  \citenamefont {Wang}, \citenamefont {Vanderbilt},\ and\ \citenamefont
  {Souza}}]{2007PRBYates}%
  \BibitemOpen
  \bibfield  {author} {\bibinfo {author} {\bibfnamefont {J.~R.}\ \bibnamefont
  {Yates}}, \bibinfo {author} {\bibfnamefont {X.}~\bibnamefont {Wang}},
  \bibinfo {author} {\bibfnamefont {D.}~\bibnamefont {Vanderbilt}},\ and\
  \bibinfo {author} {\bibfnamefont {I.}~\bibnamefont {Souza}},\ }\bibfield
  {title} {\bibinfo {title} {Spectral and {Fermi} surface properties from
  wannier interpolation},\ }\href {https://doi.org/10.1103/PhysRevB.75.195121}
  {\bibfield  {journal} {\bibinfo  {journal} {Phys. Rev. B}\ }\textbf {\bibinfo
  {volume} {75}},\ \bibinfo {pages} {195121} (\bibinfo {year}
  {2007})}\BibitemShut {NoStop}%
\bibitem [{\citenamefont {Suzuki}\ \emph {et~al.}(1992)\citenamefont {Suzuki},
  \citenamefont {Katayama}, \citenamefont {Yoshida}, \citenamefont {Tanaka},\
  and\ \citenamefont {Sato}}]{1992PRLSuzulki}%
  \BibitemOpen
  \bibfield  {author} {\bibinfo {author} {\bibfnamefont {Y.}~\bibnamefont
  {Suzuki}}, \bibinfo {author} {\bibfnamefont {T.}~\bibnamefont {Katayama}},
  \bibinfo {author} {\bibfnamefont {S.}~\bibnamefont {Yoshida}}, \bibinfo
  {author} {\bibfnamefont {K.}~\bibnamefont {Tanaka}},\ and\ \bibinfo {author}
  {\bibfnamefont {K.}~\bibnamefont {Sato}},\ }\bibfield  {title} {\bibinfo
  {title} {New magneto-optical transition in ultrathin {Fe}(100) films},\
  }\href {https://doi.org/10.1103/PhysRevLett.68.3355} {\bibfield  {journal}
  {\bibinfo  {journal} {Phys. Rev. Lett.}\ }\textbf {\bibinfo {volume} {68}},\
  \bibinfo {pages} {3355} (\bibinfo {year} {1992})}\BibitemShut {NoStop}%
\bibitem [{\citenamefont {Guo}\ and\ \citenamefont {Ebert}(1995)}]{1995PRBGuo}%
  \BibitemOpen
  \bibfield  {author} {\bibinfo {author} {\bibfnamefont {G.~Y.}\ \bibnamefont
  {Guo}}\ and\ \bibinfo {author} {\bibfnamefont {H.}~\bibnamefont {Ebert}},\
  }\bibfield  {title} {\bibinfo {title} {Band-theoretical investigation of the
  magneto-optical {Kerr} effect in {Fe} and {Co} multilayers},\ }\href
  {https://doi.org/10.1103/PhysRevB.51.12633} {\bibfield  {journal} {\bibinfo
  {journal} {Phys. Rev. B}\ }\textbf {\bibinfo {volume} {51}},\ \bibinfo
  {pages} {12633} (\bibinfo {year} {1995})}\BibitemShut {NoStop}%
\bibitem [{\citenamefont {Ravindran}\ \emph {et~al.}(1999)\citenamefont
  {Ravindran}, \citenamefont {Delin}, \citenamefont {James}, \citenamefont
  {Johansson}, \citenamefont {Wills}, \citenamefont {Ahuja},\ and\
  \citenamefont {Eriksson}}]{1999PRBRavindran}%
  \BibitemOpen
  \bibfield  {author} {\bibinfo {author} {\bibfnamefont {P.}~\bibnamefont
  {Ravindran}}, \bibinfo {author} {\bibfnamefont {A.}~\bibnamefont {Delin}},
  \bibinfo {author} {\bibfnamefont {P.}~\bibnamefont {James}}, \bibinfo
  {author} {\bibfnamefont {B.}~\bibnamefont {Johansson}}, \bibinfo {author}
  {\bibfnamefont {J.~M.}\ \bibnamefont {Wills}}, \bibinfo {author}
  {\bibfnamefont {R.}~\bibnamefont {Ahuja}},\ and\ \bibinfo {author}
  {\bibfnamefont {O.}~\bibnamefont {Eriksson}},\ }\bibfield  {title} {\bibinfo
  {title} {Magnetic, optical, and magneto-optical properties of {$\mathrm{Mn}X$
  ($X=\mathrm{As, Sb, Bi}$)} from full-potential calculations},\ }\href
  {https://doi.org/10.1103/PhysRevB.59.15680} {\bibfield  {journal} {\bibinfo
  {journal} {Phys. Rev. B}\ }\textbf {\bibinfo {volume} {59}},\ \bibinfo
  {pages} {15680} (\bibinfo {year} {1999})}\BibitemShut {NoStop}%
\bibitem [{\citenamefont {Eibschütz}\ \emph {et~al.}(1973)\citenamefont
  {Eibschütz}, \citenamefont {Sherwood}, \citenamefont {Hsu},\ and\
  \citenamefont {Cox}}]{1973AIPCPEib}%
  \BibitemOpen
  \bibfield  {author} {\bibinfo {author} {\bibfnamefont {M.}~\bibnamefont
  {Eibschütz}}, \bibinfo {author} {\bibfnamefont {R.~C.}\ \bibnamefont
  {Sherwood}}, \bibinfo {author} {\bibfnamefont {F.~S.~L.}\ \bibnamefont
  {Hsu}},\ and\ \bibinfo {author} {\bibfnamefont {C.~E.}\ \bibnamefont {Cox}},\
  }\bibfield  {title} {\bibinfo {title} {Magnetic ordering of the linear chain
  antiferromagnet {CsMnBr$_3$}},\ }\href@noop {} {\bibfield  {journal}
  {\bibinfo  {journal} {AIP Conf. Proc.}\ }\textbf {\bibinfo {volume} {10}},\
  \bibinfo {pages} {684} (\bibinfo {year} {1973})}\BibitemShut {NoStop}%
\end{thebibliography}%

\end{document}


\title{Supplementary Material for ``Time-reversal-odd Transport in Odd-Parity Magnets"}

\author{Ling Bai}
\affiliation{Key Lab of Advanced Optoelectronic Quantum Architecture and Measurement (MOE), Beijing Key Lab of Nanophotonics and Ultrafine Optoelectronic Systems, and School of Physics,Beijing Institute of Technology, Beijing 100081, China}

\author{Si Li}
\email{sili@nwu.edu.cn}
\affiliation{School of Physics, Northwest University, Xi'an 710127, China}

\author{Libor \v{S}mejkal}
\affiliation{Institute of Physics, Johannes Gutenberg University Mainz, 55099 Mainz, Germany}
\affiliation{Institute of Physics, Czech Academy of Sciences, Cukrovarnická 10, 162 00 Praha 6, Czech Republic}

\author{Yugui Yao}
\affiliation{Key Lab of Advanced Optoelectronic Quantum Architecture and Measurement (MOE), Beijing Key Lab of Nanophotonics and Ultrafine Optoelectronic Systems, and School of Physics,Beijing Institute of Technology, Beijing 100081, China}

\author{Wanxiang Feng}
\email{wxfeng@bit.edu.cn}
\affiliation{Key Lab of Advanced Optoelectronic Quantum Architecture and Measurement (MOE), Beijing Key Lab of Nanophotonics and Ultrafine Optoelectronic Systems, and School of Physics,Beijing Institute of Technology, Beijing 100081, China}

\date{\today}

\maketitle

\section{I. DERIVATION OF THE SOC-FREE HAMILTONIAN $H_0(\mathbf{k})$}
In this section, we provide the details of the theoretical models introduced in the main text.
We consider the enlarged $\sqrt{3}\times\sqrt{3}$ magnetic unit cell, whose primitive lattice vectors $\mathbf{a}_1$ and $\mathbf{a}_2$ are given by:
\begin{equation}
	\mathbf{a}_1 = a (1, 0, 0), \quad \mathbf{a}_2 = a ( -\frac{1}{2}, \frac{\sqrt{3}}{2}, 0).
\end{equation}
Each magnetic unit cell hosts six sublattices denoted as $\alpha \in \{A_t, B_t, C_t, A_b, B_b, C_b\}$, where the subscripts $t$ and $b$ represent the top and bottom layers, respectively. Within each layer, every sublattice site has nearest neighbors belonging to the other two sublattices. Let us define the three primary nearest-neighbor vectors connecting adjacent sites in the original triangular lattice:
\begin{equation}
	\mathbf{d}_1 = a ( 0, -\frac{\sqrt{3}}{3}, 0), \quad \mathbf{d}_2 = a (-\frac{1}{2}, \frac{\sqrt{3}}{6}, 0), \quad \mathbf{d}_3 = a (\frac{1}{2}, \frac{\sqrt{3}}{6}, 0).
\end{equation}

The Hamiltonian without SOC is given by:
\begin{align}
	{H}_0=& \sum_{\langle ij\rangle,\sigma} t \, {c}_{i\sigma}^\dagger \, {c}_{j\sigma}+\sum_{\langle ij\rangle,\sigma} t_{\perp} \, {c}_{i\sigma}^\dagger \, {c}_{j\sigma} + J \sum_{i, \sigma, \sigma'} \mathbf{m}_i \cdot c_{i \sigma}^\dagger \boldsymbol{\sigma}_{\sigma \sigma'} c_{i \sigma'}\notag,
\end{align}

To perform the spatial Fourier transform, the electron annihilation operator at site $i$ of sublattice $\alpha$ with spin $\sigma \in \{\uparrow, \downarrow\}$ is defined as:
\begin{equation}
	c_{i\alpha\sigma} = \frac{1}{\sqrt{N}} \sum_{\mathbf{k}} e^{i \mathbf{k} \cdot \mathbf{R}_{i\alpha}} c_{\alpha\sigma}(\mathbf{k}),
\end{equation}
where $\mathbf{R}_{i\alpha} = \mathbf{R}_i + \mathbf{r}_\alpha$, and $\mathbf{R}_i = n_1 \mathbf{a}_1 + n_2 \mathbf{a}_2$ labels the lattice vector of the $i$-th magnetic unit cell ($n_1, n_2 \in \mathbb{Z}$). The full 12-dimensional Nambu basis is organized as:
\begin{equation}
	\Psi(\mathbf{k}) = \Big( c_{A_t\uparrow}, c_{A_t\downarrow}, c_{B_t\uparrow}, c_{B_t\downarrow}, c_{C_t\uparrow}, c_{C_t\downarrow}, c_{A_b\uparrow}, c_{A_b\downarrow}, c_{B_b\uparrow}, c_{B_b\downarrow}, c_{C_b\uparrow}, c_{C_b\downarrow} \Big)^T.
\end{equation}
In this representation, the total Hamiltonian matrix $H_0(\mathbf{k})$ satisfies the following layer-blocked structure:
\begin{equation}
	H_0(\mathbf{k}) = \begin{pmatrix}
		H_{\text{top}}(\mathbf{k}) & H_{\perp}(\mathbf{k}) \\
		H_{\perp}^\dagger(\mathbf{k}) & H_{\text{bottom}}(\mathbf{k})
	\end{pmatrix}.
\end{equation}
By performing the spatial Fourier transform with respect to the real-space hopping paths connecting the sublattices $A$, $B$, and $C$, the kinetic hopping phase factors can be grouped into three complex structural functions:
\begin{align}
	\gamma_{AB}(\mathbf{k}) &= e^{i\mathbf{k}\cdot\mathbf{d}_1} + e^{i\mathbf{k}\cdot\mathbf{d}_2} + e^{i\mathbf{k}\cdot\mathbf{d}_3}, \\
	\gamma_{BC}(\mathbf{k}) &= \gamma_{AB}(\mathbf{k}), \\
	\gamma_{CA}(\mathbf{k}) &= \gamma_{AB}(\mathbf{k}).
\end{align}

The exchange term models the coupling between itinerant electron spins and local coplanar 120$^\circ$ magnetic moments $\mathbf{m}_\alpha$. Utilizing the standard Pauli matrices $\boldsymbol{\sigma} = (\sigma_x, \sigma_y, \sigma_z)$, the exchange potentials $V_\alpha = J \mathbf{m}_\alpha \cdot \boldsymbol{\sigma}$ are evaluated.
For the top layer, substituting $\mathbf{m}_{A_t} = (-1/2, -\sqrt{3}/2, 0)$, $\mathbf{m}_{B_t} = (1, 0, 0)$, and $\mathbf{m}_{C_t} = (-1/2, \sqrt{3}/2, 0)$ gives:
\begin{align}
	V_{A_t} &= J \left( -\frac{1}{2}\sigma_x - \frac{\sqrt{3}}{2}\sigma_y \right) = \frac{J}{2} \begin{pmatrix} 0 & -1+i\sqrt{3} \\ -1-i\sqrt{3} & 0 \end{pmatrix}, \\
	V_{B_t} &= J \sigma_x = J \begin{pmatrix} 0 & 1 \\ 1 & 0 \end{pmatrix}, \\
	V_{C_t} &= J \left( -\frac{1}{2}\sigma_x + \frac{\sqrt{3}}{2}\sigma_y \right) = \frac{J}{2} \begin{pmatrix} 0 & -1-i\sqrt{3} \\ -1+i\sqrt{3} & 0 \end{pmatrix}.
\end{align}
For the bottom layer, the magnetic moments are strictly reversed ($\mathbf{m}_{A_b} = -\mathbf{m}_{A_t}$, etc.), yielding $V_{\alpha_b} = -V_{\alpha_t}$.

Combining the hopping and exchange terms, the $6\times6$ diagonal layer blocks of $H_0(\mathbf{k})$ are:
\begin{align}
	H_{\text{top}}(\mathbf{k}) &= \begin{pmatrix}
		V_{A_t} & t\gamma_{AB}(\mathbf{k})\sigma_0 & t\gamma_{CA}^*(\mathbf{k})\sigma_0 \\
		t\gamma_{AB}^*(\mathbf{k})\sigma_0 & V_{B_t} & t\gamma_{BC}(\mathbf{k})\sigma_0 \\
		t\gamma_{CA}(\mathbf{k})\sigma_0 & t\gamma_{BC}^*(\mathbf{k})\sigma_0 & V_{C_t}
	\end{pmatrix}, \\
	H_{\text{bottom}}(\mathbf{k}) &= \begin{pmatrix}
		-V_{A_t} & t\gamma_{AB}(\mathbf{k})\sigma_0 & t\gamma_{CA}^*(\mathbf{k})\sigma_0 \\
		t\gamma_{AB}^*(\mathbf{k})\sigma_0 & -V_{B_t} & t\gamma_{BC}(\mathbf{k})\sigma_0 \\
		t\gamma_{CA}(\mathbf{k})\sigma_0 & t\gamma_{BC}^*(\mathbf{k})\sigma_0 & -V_{C_t}
	\end{pmatrix},
\end{align}
where $\sigma_0$ is the $2\times2$ identity matrix. The off-diagonal block for $H_0(\mathbf{k})$ is $H_{\perp} = t_\perp \mathbb{I}_{6\times6}$ due to the spin-conserving vertical interlayer hopping.

\section{II. DERIVATION OF THE SOC-ACTIVE HAMILTONIAN $H_1(\mathbf{k})$}

The total Hamiltonian can be written as
\begin{align}
	H = H_0 + H_1,
	\label{eq:Htot}
\end{align}
where $H_0$ is the SOC-free term and $H_1$ collects the SOC-related contributions. In our minimal model,
\begin{align}
	H_1
	=&\sum_{\langle ij\rangle,\sigma,\sigma'} r \, c_{i\sigma}^\dagger \, c_{j\sigma'}
	+ i\lambda_R \sum_{\langle ij\rangle} c_i^\dagger \left[ (\boldsymbol{\sigma} \times \mathbf{d}_{ij}) \cdot \hat{\mathbf{z}} \right] c_j,
	\label{eq:H1_SM}
\end{align}
where the first term describes a spin-mixing interlayer hopping with amplitude $r$, and the second term is the interface Rashba SOC with coupling strength $\lambda_R$, the in-plane unit vector $\mathbf{d}_{ij}$ pointing from site $j$ to site $i$, and the out-of-plane unit normal vector $\hat{\mathbf{z}}$. The Rashba SOC arises from inversion-symmetry breaking at the bilayer interface. The spin-mixing interlayer hopping modifies the interlayer coupling block of total Hamiltonian, i.e., $H_{\perp}$, which is obtained by using the MagneticTB package~\cite{zhang2022magnetictb}:
\begin{equation}
	H_{\perp} =
	\begin{pmatrix}
		t_{\perp} & (-1+i\sqrt{3})r & 0 & 0 & 0 & 0 \\
		(1 + i\sqrt{3})r & t_{\perp} & 0 & 0 & 0 & 0 \\
		0 & 0 & t_{\perp} & 2r & 0 & 0 \\
		0 & 0 & -2r & t_{\perp} & 0 & 0 \\
		0 & 0 & 0 & 0 & t_{\perp} & (-1 - i\sqrt{3})r \\
		0 & 0 & 0 & 0 & (1 - i\sqrt{3})r & t_{\perp}
	\end{pmatrix}.
	\label{eq:Hperp_SM}
\end{equation}
\newpage

\section{III. DETAILS OF FIRST-PRINCIPLES CALCULATIONS}
First-principles calculations were performed using the Vienna \textit{ab initio} Simulation Package (VASP)~\cite{VASP} within the projector augmented-wave (PAW) method~\cite{PAWa,PAWb}. A plane-wave cutoff energy of 500 eV was used. The generalized gradient approximation (GGA)\cite{GGA} of Perdew–Burke–Ernzerhof (PBE)\cite{PBE} was opted to treat the exchange-correlation function. All atoms were fully relaxed with a self-consistency energy criterion of 10$^{-6}$ eV until the Hellmann–Feynman forces on each atom were less than 10$^{-3}$ eV/\AA. The Brillouin zone was sampled using a $\Gamma$-centered $7 \times  7 \times 1$ $k$-mesh. The localized $d$ orbitals of Fe atoms were treated with $U = 4$ eV. A vacuum layer of more than 20~\AA\ was introduced along the out-of-plane direction to avoid spurious interactions between periodic slabs. The DFT-D3 van der Waals correction was adopted to describe the interlayer interaction~\cite{DFT-D3}. Effective Hamiltonians based on maximally localized Wannier functions were constructed using the \textsc{WANNIER90} package~\cite{Wannier90}. For the Fe-C-Fe heterostructure, 240 Wannier functions were adopted, including Fe-$d$, C-$p$, and O-$p$ orbitals from the Fe$_3$C$_6$O$_6$ layers and C-$p_z$ orbitals from the graphene layer.

On the basis of Wannier functions, the intrinsic anomalous Hall conductivity was evaluated on a dense $k$-mesh of $1000 \times 1000 \times 1$ using the Kubo formula~\cite{2004PRLYao}, 
\begin{equation}
	\begin{aligned}
		\sigma_{x y}(\varepsilon)=\frac{e^2}{\hbar} \sum_n \int \frac{d^2 k}{(2 \pi)^2} \Omega^n_{x y}(k)f_n(k),
	\end{aligned}
\end{equation}
where $f_n(k)$ is the Fermi-Dirac distribution function, and $\Omega^n_{x y}(k)$ is the band- and momentum-resolved Berry curvature, given by
\begin{equation}
	\begin{aligned}
		\Omega_{x y}^n(k)=-\sum_{n^{\prime} \neq n} \frac{2 \operatorname{Im}[\langle\psi_{n k}| \hat{v}_x |\psi_{n^{\prime} k}\rangle \langle\psi_{n^{\prime} k}| \hat{v}_y | \psi_{n k}\rangle]}{\left(\varepsilon_{n k}-\varepsilon_{n^{\prime} k}\right)^2},
	\end{aligned}
\end{equation}
where $\hat{v}_{x/y}$ are the velocity operators and $\psi_{n k}$ ($\varepsilon_{n k}$) is the eigenvector (eigenvalue) at band index $n$ and momentum $k$.

The optical conductivity $\sigma_{ij}(\omega)$ can be calculated by using the Kubo-Greenwood formula~\cite{2007PRBYates},
\begin{align}
	\sigma_{ij}(\omega)
	&= \sigma_{ij}^{\prime}(\omega)+i\sigma_{ij}^{\prime\prime}(\omega)  \notag \\
	&= \frac{i e^{2}\hbar}{N_k V}
	\sum_{\bm{k}}\sum_{n,n^\prime}
	\frac{f_{n^\prime\bm{k}}-f_{n\bm{k}}}{\varepsilon_{n^\prime\bm{k}}-\varepsilon_{n\bm{k}}}
	\frac{\left\langle \psi_{n\bm{k}} \middle| \hat{v}_i \middle| \psi_{n^\prime\bm{k}} \right\rangle
		\left\langle \psi_{n^\prime\bm{k}} \middle| \hat{v}_j \middle| \psi_{n\bm{k}} \right\rangle}
	{\varepsilon_{n^\prime\bm{k}}-\varepsilon_{n\bm{k}}-(\hbar\omega+i\eta)}, 
\end{align}
here, the subscripts $i,j\in \{x,y\}$ denote Cartesian coordinates, and $\sigma_{ij}^{\prime}(\omega)$ and $\sigma_{ij}^{\prime\prime}(\omega)$ represent the real and imaginary parts of $\sigma_{ij}(\omega)$ at frequency of incident light $\omega$, respectively. $V$ is the cell volume, and $N_k$ is the total number of $k$ points used to sample the Brillouin zone. In the present calculations, a $k$-mesh of $700 \times 700 \times 1$ was used. The quantity $\hbar \omega$ is the photon energy, and $\eta$ is an adjustable broadening parameter for energy smearing.

The complex Kerr angle $\phi_K$ and Faraday angle $\phi_F$ are given by~\cite{1992PRLSuzulki,1995PRBGuo,1999PRBRavindran}
\begin{align}
	&\phi_K=\theta_K+i\eta_K=-\frac{8\pi d\,\sigma_{xy}(\omega)}{c\left(1-n_{\mathrm{sub}}^{2}\right)}, \\
	&\phi_F=\theta_F+i\eta_F=
	\frac{2\pi d}{c}\cdot
	\frac{\sigma_{xy}(\omega)}
	{\sqrt{1+\dfrac{4\pi i}{\omega}\sigma_{xx}(\omega)}},
\end{align}
where $\theta_{K/F}$ and $\eta_{K/F}$ denote the Kerr/Faraday rotation angle and ellipticity, respectively. The quantities $c$, $d$, and $n_{\mathrm{sub}}$ denote the speed of light in vacuum, the film thickness, and the refractive index of the nonmagnetic substrate, respectively.

\newpage

\section{IV. 3D MATERIAL CANDIDATES}
\begin{figure}[htbp]
	\centering
	\includegraphics[width=1\columnwidth]{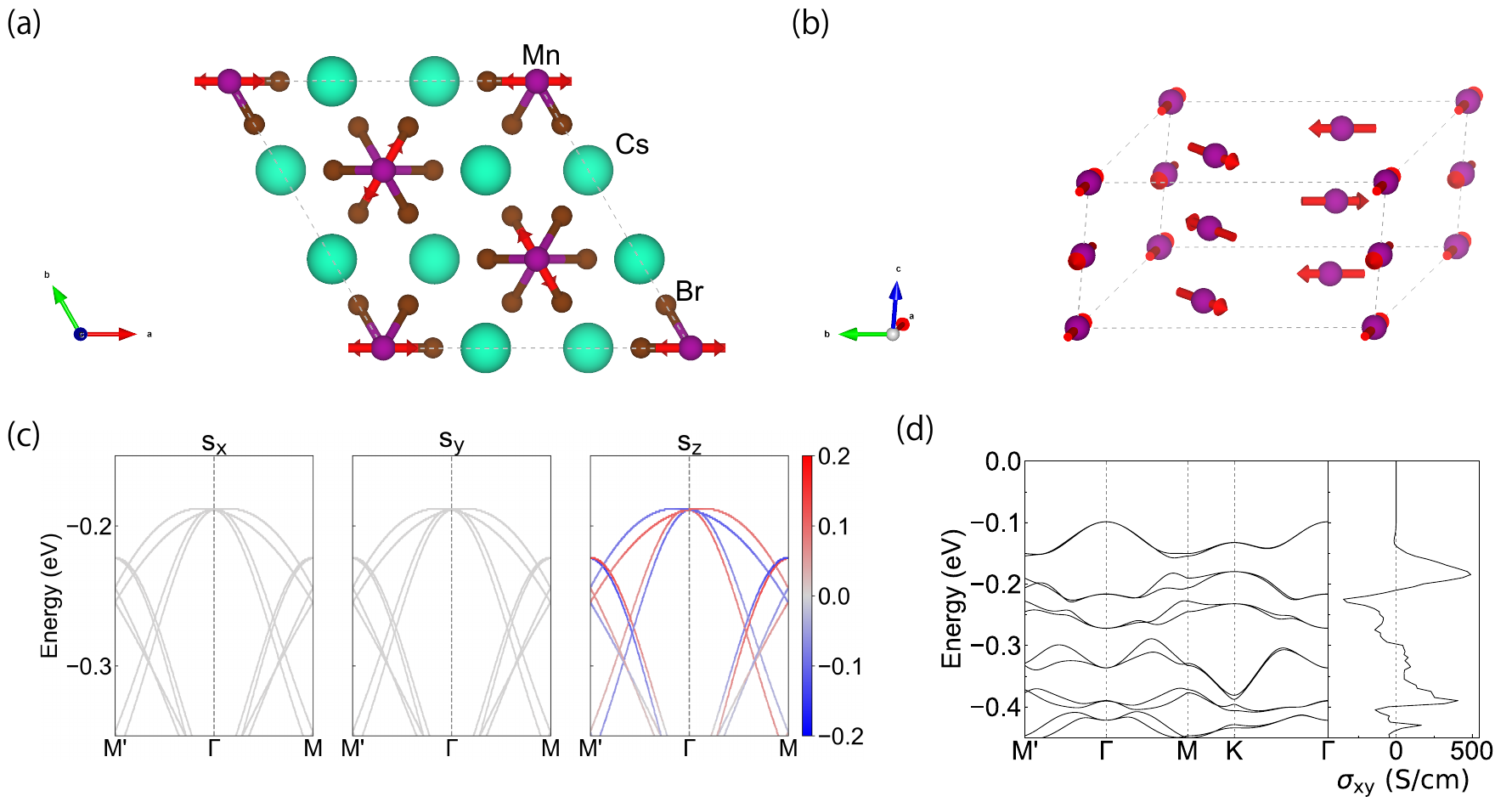}
	\caption{Crystal, magnetic, and electronic structures of CsMnBr$_3$ (a) Top view. (b) Crystal structure with only magnetic atoms. The red arrows in (a,b) indicate the magnetic moments on Mn atoms.  (c) Spin-projected nonrelativistic band structure, with odd-parity spin splitting, i.e., $s_{z}(\mathbf{k}) = -s_{z}(-\mathbf{k})$ and $s_{x,y}(\mathbf{k})=0$. (d) Relativistic band structure and anomalous Hall conductivity $\sigma_{xy}$.}
	\label{figS1}
\end{figure}

By combining MAGNDATA magnetic materials database~\cite{2016ACGallego} with FINDSPINGROUP program~\cite{2026arXivYu}, we identify experimentally synthesized coplanar noncollinear antiferromagnets as candidate odd-parity magnets that can host time-reversal-odd transport, as listed in Table~\ref{tab:TabS1}. In particular, we determine whether anomalous Hall-type responses are symmetry allowed for each candidate. For Ba$_3$MnNb$_2$O$_9$ (1.0.8) and Ba$_3$CoSb$_2$O$_9$ (1.0.45), the magnetic configurations reported in MAGNDATA forbid the AHE, whereas an in-plane rotation of the magnetic moments can render the AHE symmetry allowed. In contrast, for CsFeCl$_3$ (1.0.14) and ThMn$_2$ (1.0.24), the AHE remains symmetry forbidden for any in-plane orientation of the magnetic moments. Nevertheless, these magnetic states can support other types of time-reversal-odd responses, such as those associated with the quantum metric dipole. Here, taking CsMnBr$_3$ (1.0.35) as a representative material~\cite{1973AIPCPEib}, we present its band structures without and with spin-orbit coupling, together with the corresponding anomalous Hall conductivity as shown in Fig~\ref{figS1}.

\begin{table}[htbp]
	\renewcommand{\arraystretch}{1.5}
	\centering
	\caption{List of 3D material candidates, together with their MAGNDATA IDs, spin space groups, magnetic space groups, even-wave symmetries, and symmetry-allowance of anomalous Hall-type responses.}
	\label{tab:TabS1}
	\begin{ruledtabular}
 		\begin{tabular}{cccccc}
		Material ID& Formula &Spin space group & Magnetic space group &Odd-parity type & AHE  \\ 
		\hline
        1.0.8  & Ba$_3$MnNb$_2$O$_9$ & 164.149.3.1.P & 157.53 & \textit{f}-wave &\ding{55}\\
        1.0.14 & CsFeCl$_3$ & 194.188.3.1.P & 189.223 & \textit{f}-wave &\ding{55}\\
        1.0.24 & ThMn$_2$ & 194.174.3.1.P & 189.223 & \textit{f}-wave &\ding{55}\\
        1.0.32 & EuIn$_2$As$_2$ & 194.156.3.1.P & 178.159 & \textit{p}-wave &\ding{51}\\
		1.0.35 & CsMnBr$_3$ & 194.149.3.1.P & 189.225 & \textit{f}-wave& \ding{51}\\
        1.0.40 & RbFeCl$_3$ & 194.188.3.1.P & 189.224 & \textit{f}-wave& \ding{51}\\
        1.0.41 & RbNiCl$_3$ & 194.149.3.1.P & 20.34 & \textit{f}-wave& \ding{51}\\
        1.0.42 & CsNiCl$_3$ & 194.149.3.1.P & 20.34 & \textit{f}-wave& \ding{51}\\
        1.0.44 & Ba$_3$CoSb$_2$O$_9$ & 194.149.3.1.P & 36.174 & \textit{f}-wave& \ding{51}\\
        1.0.45 & Ba$_3$CoSb$_2$O$_9$ & 194.149.3.1.P & 189.221 & \textit{f}-wave& \ding{55}\\
        1.0.46 & Ba$_3$MnSb$_2$O$_9$ & 15.1.3.7.P & 5.13 & \textit{p}-wave& \ding{51}\\
		\end{tabular}
	\end{ruledtabular}
\end{table}

\clearpage

\section{V. SUPPLEMENTARY FIGURES AND TABLE}

\begin{figure}[htbp]
	\centering
	\includegraphics[width=0.9\columnwidth]{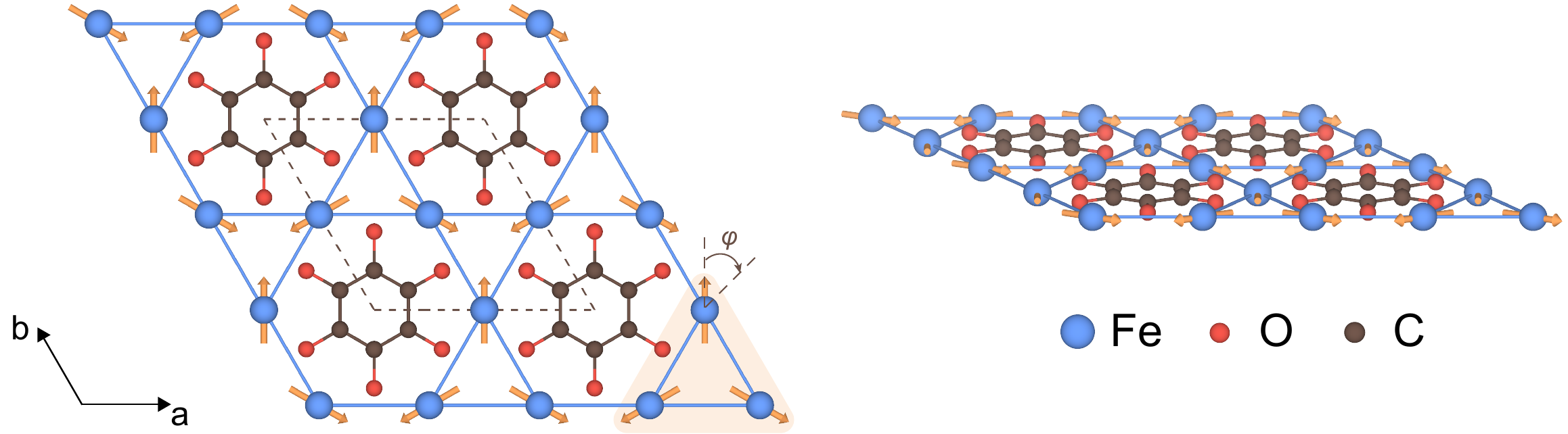}
	\caption{Crystal and magnetic structures of monolayer Fe$_3$C$_6$O$_6$. Gold arrows denote the spin magnetic moments on Fe atoms, forming a $120^\circ$ coplanar noncollinear configuration. The angle $\varphi$ parameterizes a uniform rotation of the three spins away from their reference directions (radially outward from the center of the triangle).}
	\label{figS2}
\end{figure}

\begin{figure}[htbp]
	\centering
	\includegraphics[width=0.9\columnwidth]{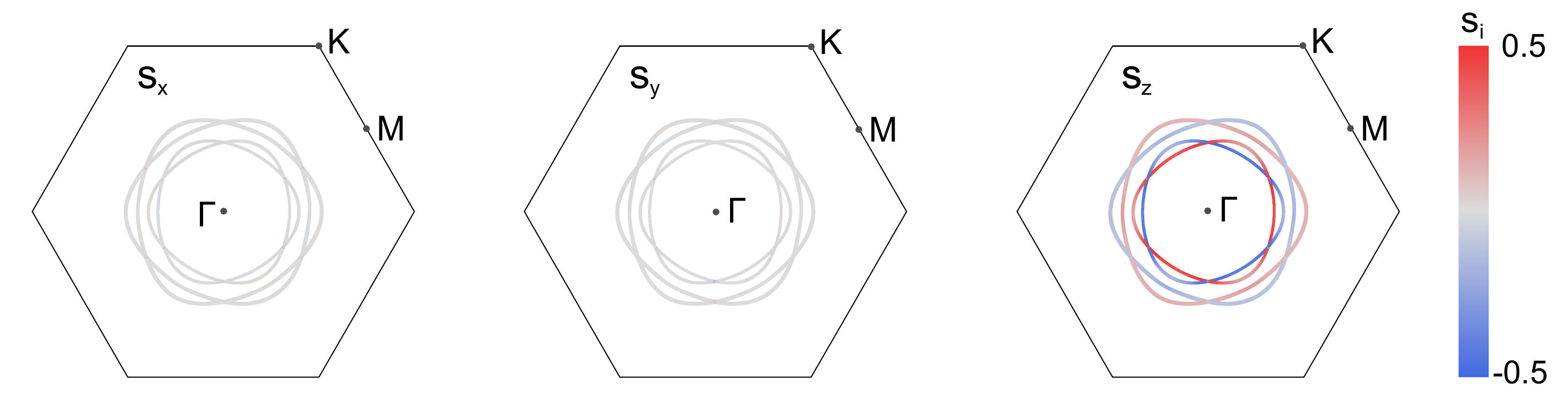}
	\caption{Spin-projected nonrelativistic Fermi surface at $E=E_\mathrm{VBM}-1.04$ eV of the Fe-C-Fe heterostructure for $\varphi=0^\circ$. The corresponding spin space group is $p^{6^1_{001}}\bar{6}^{2_{120}}m^{2_{010}}2^{m_{001}}1$, whose nontrivial spin space group operations include $[E\,\|\, E]$, $[C_{2[210]}\|\, M_{[010]}]$, $[C_{2[010]}\,\|\, C_{2[210]}]$, $[C_{2z}\,\|\, M_z]$, $[C_{2[120]}\,\|\, M_{[100]}]$, $[C_{2[100]}\,\|\, C_{2[120]}]$, $[C_{2[1\bar{1}0]}\,\|\, M_{[110]}]$, $[C_{2[110]}\,\|\, C_{2[1\bar{1}0]}]$, $[C^{+}_{3z}\,\|\, C^{+}_{3z}]$, $[C^{-}_{6z}\,\|\, \bar{C}^{-}_{6z}]$, $[C^{-}_{3z}\,\|\, C^{-}_{3z}]$, and $[C^{+}_{6z}\,\|\, \bar{C}^{+}_{6z}]$.}
	\label{figS3}
\end{figure}

\begin{figure}[htbp]
	\centering
	\includegraphics[width=1\columnwidth]{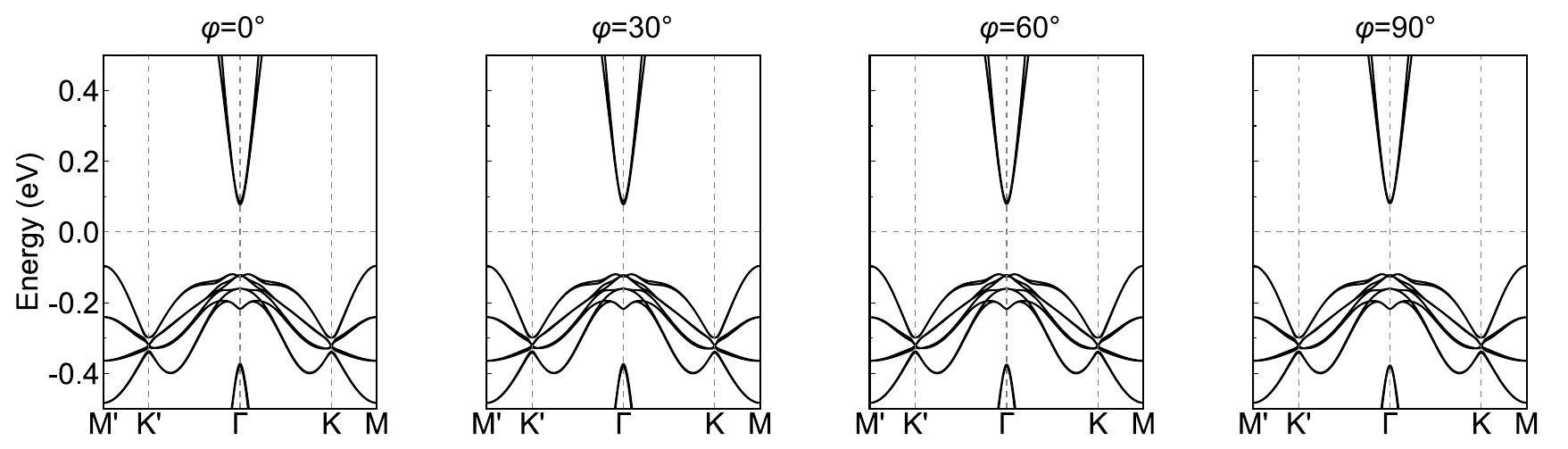}
	\caption{Relativistic band structure of the Fe-C-Fe heterostructure for $\varphi=0^\circ$, $30^\circ$, $60^\circ$, and $90^\circ$.}
	\label{figS4}
\end{figure}

\begin{figure}[htbp]
	\centering
	\includegraphics[width=0.6\columnwidth]{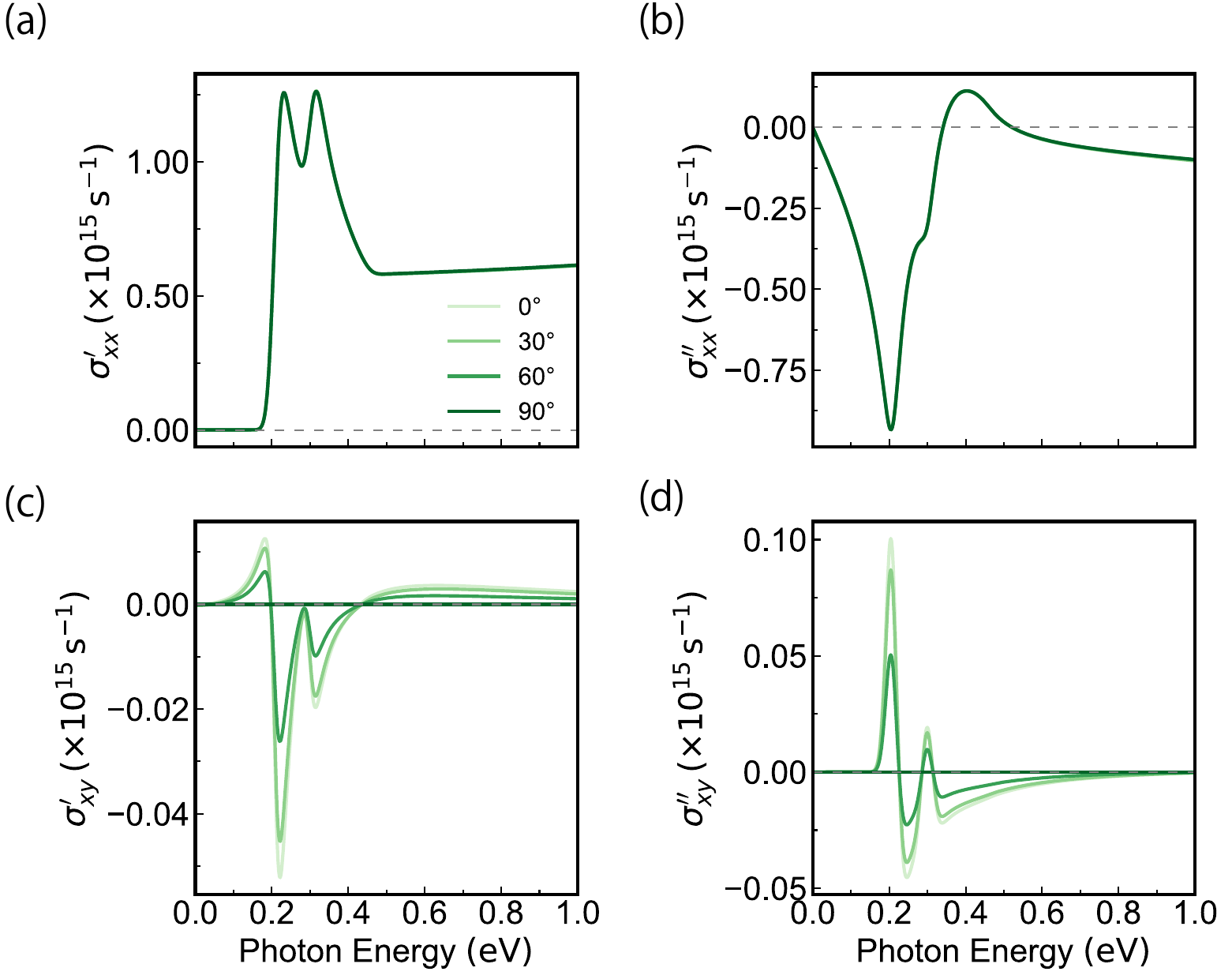}
	\caption{(a) Real and (b) imaginary parts of the diagonal optical conductivity $\sigma_{xx}(\omega)$, and (c) real and (d) imaginary parts of the off-diagonal optical conductivity $\sigma_{xy}(\omega)$ of the Fe–C–Fe heterostructure for $\varphi=0^\circ$, $30^\circ$, $60^\circ$, and $90^\circ$.}
	\label{figS5}
\end{figure}

\begin{table}[htbp]
	\renewcommand{\arraystretch}{1.5}
	\centering
	\caption{Spin space groups, magnetic space groups, types of odd-parity spin-momentum locking, and symmetry-allowance of anomalous Hall-type responses in the Fe-C-Fe heterostructure for different spin-orientation angles $\varphi$.}
	\label{tab:TabS2}
	\begin{ruledtabular}
 		\begin{tabular}{ccccc}
		Azimuthal angle& Spin space group&Magnetic space group &Odd-parity type& AHE  \\ 
		\hline
		0$^\circ$ &$p^{6^1_{001}}\bar{6}^{2_{120}}m^{2_{010}}2^{m_{001}}1$&$p\bar{6}m^\prime 2^\prime$ &\textit{f}-wave&\ding{51}\\
		15$^\circ$&$p^{6^1_{001}}\bar{6}^{2_{100}}m^{2_{210}}2^{m_{001}}1$&$p\bar{6}$&\textit{f}-wave&\ding{51}\\
		30$^\circ$&$p^{6^1_{001}}\bar{6}^{2_{110}}m^{2_{120}}2^{m_{001}}1$&$p\bar{6}$&\textit{f}-wave&\ding{51}\\
		45$^\circ$&$p^{6^1_{001}}\bar{6}^{2_{100}}m^{2_{210}}2^{m_{001}}1$&$p\bar{6}$&\textit{f}-wave&\ding{51}\\
		60$^\circ$&$p^{6^1_{001}}\bar{6}^{2_{210}}m^{2_{110}}2^{m_{001}}1$&$p\bar{6}$&\textit{f}-wave&\ding{51}\\
		75$^\circ$&$p^{6^1_{001}}\bar{6}^{2_{100}}m^{2_{210}}2^{m_{001}}1$&$p\bar{6}$&\textit{f}-wave&\ding{51}\\
		90$^\circ$&$p^{6^1_{001}}\bar{6}^{2_{100}}m^{2_{210}}2^{m_{001}}1$&$p\bar{6}m2$&\textit{f}-wave&\ding{55}\\
		105$^\circ$&$p^{6^1_{001}}\bar{6}^{2_{100}}m^{2_{210}}2^{m_{001}}1$&$p\bar{6}$&\textit{f}-wave&\ding{51}\\
		120$^\circ$&$p^{6^1_{001}}\bar{6}^{2_{\bar{1}10}}m^{2_{100}}2^{m_{001}}1$&$p\bar{6}$&\textit{f}-wave&\ding{51}\\
		135$^\circ$&$p^{6^1_{001}}\bar{6}^{2_{100}}m^{2_{210}}2^{m_{001}}1$&$p\bar{6}$&\textit{f}-wave&\ding{51}\\
		150$^\circ$&$p^{6^1_{001}}\bar{6}^{2_{010}}m^{2_{\bar{1}10}}2^{m_{001}}1$&$p\bar{6}$&\textit{f}-wave&\ding{51}\\
		165$^\circ$&$p^{6^1_{001}}\bar{6}^{2_{100}}m^{2_{210}}2^{m_{001}}1$&$p\bar{6}$&\textit{f}-wave&\ding{51}\\
		180$^\circ$&$p^{6^1_{001}}\bar{6}^{2_{120}}m^{2_{010}}2^{m_{001}}1$&$p\bar{6}m^\prime 2^\prime$ &\textit{f}-wave&\ding{51}\\
			\end{tabular}
	\end{ruledtabular}
\end{table}

\clearpage

\bibliography{mybib.bib}